\documentclass[final,authoryear,1p, times, twocolumn]{elsarticle}
\usepackage{graphicx}
\usepackage{amssymb}
\usepackage{natbib}
\usepackage{url}

\usepackage{multirow}
\usepackage{color}	
\RequirePackage{rotating}

\journal{New Astronomy}

\newcommand{\db}[1]{\texttt{#1}} 	
\newcommand{\dbtab}[1]{\texttt{#1}} 
\newcommand{\dbcol}[1]{\emph{#1}} 

\begin{document}
\begin{frontmatter}

\title{The MultiDark Database: Release of the Bolshoi and MultiDark Cosmological Simulations}
\author[aip]{Riebe, Kristin}
\ead{kriebe@aip.de}
\author[aip]{Partl, Adrian M.}
\ead{apartl@aip.de}
\author[aip]{Enke, Harry}
\ead{henke@aip.de}
\author[aip]{Forero-Romero, Jaime}
\ead{jforero@aip.de}
\author[aip]{Gottl\"ober, Stefan}
\ead{sgottloeber@aip.de}
\author[nmsu]{Klypin, Anatoly}
\ead{aklypin@nmsu.edu}
\author[mpa]{Lemson, Gerard}
\ead{lemson@mpa-garching.mpg}
\author[iaa]{Prada, Francisco}
\ead{fprada@iaa.es}
\author[ucsc]{Primack, Joel R.}
\ead{joel@ucsc.edu}
\author[aip]{Steinmetz, Matthias}
\ead{msteinmetz@aip.de}
\author[IAM]{Turchaninov, Victor}
\ead{vturch@utech.ru}
\address[aip]{Leibniz-Institut f\"ur Astrophysik Potsdam (AIP), An der Sternwarte 16, 14482 Potsdam (Germany)}
\address[nmsu]{Astronomy Department, New Mexico State University (NMSU), Las Cruces, NM 88001 (USA)}
\address[mpa]{Max Planck Institut f\"ur Astrophysik (MPA), Karl-Schwarzschild-Str. 1, 85741 Garching (Germany)}
\address[iaa]{Instituto de Astrofisica de Andalucia (CSIC), Glorieta de la Astronomia S/N, 18008 Granada (Spain)}
\address[IAM]{Institute of Applied Mathematics, Miusskaya Sq.4, 125047, Moscow (Russia)}
\address[ucsc]{Department of Physics, University of California, Santa Cruz, CA 95064 (USA)}

\begin{abstract}
  We present the online {\it MultiDark Database} -- a Virtual
  Observatory-oriented, relational database for hosting various cosmological simulations. 
  The data is accessible via an SQL (Structured Query Language) query interface,  
  which also allows users to directly pose scientific questions, 
  as shown in a number of examples in this paper. Further examples for the usage of the database 
  are given in its extensive online documentation.
  The database is based on the same technology 
  as the Millennium Database, a fact that will greatly facilitate the
  usage of both suites of cosmological simulations. 
  The first release of the {\it MultiDark Database} hosts 
  two 8.6~billion particle
  cosmological $N$-body simulations: the Bolshoi (250\,$h^{-1}$Mpc
  simulation box, 1\,$h^{-1}$kpc resolution) and MultiDark Run1 simulation (MDR1, or
  BigBolshoi, 1000\,$h^{-1}$Mpc simulation box, 7\,$h^{-1}$kpc
  resolution). 
  The extraction
  methods for halos/subhalos from the raw simulation data, and how
  this data is structured in the database are explained in this paper. 
  With the first data release, users get full access to halo/subhalo catalogs,
  various profiles of the halos at redshifts $z=0-15$, and raw dark
  matter data for one time-step of the Bolshoi and four time-steps of the
  MultiDark simulation. Later releases will also include galaxy mock
  catalogs and additional merging trees for both simulations as well as new large volume
  simulations with high resolution.  
  This project is further proof of the viability to store and present
  complex data using relational database technology. We encourage other simulators
  to publish their results in a similar manner.

\end{abstract}


\end{frontmatter}
\section{Introduction}
\label{sec:intro}

Computer simulations play a very important role in cosmology. The
field started in the 1960s and 1970s with $N-$body
simulations which had just a few hundred or thousand particles 
\citep{Aarseth1966, Aarseth1969, Peebles1970, White1976, Gott1979, Efstathiou1979}. 
Thanks to the steady improvement of computer hardware and computational algorithms,
we now have large simulations with many billions of particles
\citep{millennium,2009MNRAS.398.1150B, Teyssier2009, Kim2009,
  Klypin.Trujillo-Gomez.Primack.2010,Wetzel2010, Prada2011,Iliev2011}. These and other large
  $N-$body simulations are used to address numerous aspects of the 
  evolution of fluctuations and formation of dark matter halos. They
  provide extraordinary accuracy for such important statistics as the
  mass function of halos \citep[e.g.,][]{Jenkins2001, Sheth2002, Warren2006, Tinker2008,
Klypin.Trujillo-Gomez.Primack.2010}, halo concentration \citep{Bullock2001, Zhao2003, Neto2007, Maccio2008,
Prada2011, Iliev2011}, halo  correlation
  function and biases \citep{Jing1998, Kravtsov1999, Gao2005, Wechsler2006, Tinker2010}, 
statistics and distribution of satellites \citep{Klypin1999, Moore1999, Springel2008, Kuhlen2008}, 
and dark matter density profiles \citep{Dubinski1991, NFW1997, Springel2008, Stadel2009}.

In spite of the fact that cosmological $N-$body simulations give information only on
dark matter and do not mimic the evolution of baryons, there are
different ways to make theoretical predictions for ``galaxies'' in
these dark-matter-only simulations. For instance, one can use semi-analytical
methods to predict properties of galaxies hosted by dark matter halos
and subhalos
\citep[e.g.,][]{Kauffmann1999,Somerville1999,Croton2006,Somerville2011}. Another
option is to use the Halo-Occupation-Distribution 
\citep[HOD;][]{Kravtsov2004,Vale2004,Zentner2005,Bosch2007} to split halos
into ``galaxies''. A third option is Halo-Abundance-Matching
 \citep[HAM;][]{Kravtsov2004,Conroy2006,Trujillo2010} by matching the largest halos
(and subhalos) with the brightest galaxies.

However, with the growing size of numerical simulation data some
problems evolved. The sheer amount of data in large simulations is
difficult to handle. The simulations typically provide so much data
that even a compact form of different ``catalogs'' may become
impractical to distribute to all people involved in different
research groups, not to mention to provide it to the whole
astronomical community, or to release raw simulation data.
Accessing data written in different formats puts an extra burden on
people who intend to analyze the results of the simulations.

There are different ways of handling the situation. We decided to use
a database, which, in combination with its powerful {\it Structured
  Query Language} (SQL), allows users to filter the data on the server
side, analyze the resulting subsets of the data, and retrieve only their results.
Since the amount of data these simulations produce lies nowadays
in the multi-terabyte range, the full data set is generally too
large to retrieve and manage for most users. Server side filtering is
a prerequisite for successful dissemination of such large data sets.

Large observational surveys like the SDSS have pioneered this approach
in astronomy and have shown that providing data directly through SQL
is a very fruitful approach \citep[][]{SDSS_Database}. The {\it Millennium
Database} \citep{2006astro.ph..8019L} played an important role by
making the Millennium simulations \citep[][]{Springel.2005,2009MNRAS.398.1150B} accessible
to numerous users. The {\it MultiDark Database} uses the same
technology, implementation and data structures as the {\it Millennium Database}, a fact that will greatly
facilitate the usage of both databases to study consistency of dark
matter halo statistics in simulations performed using different codes,
numerical algorithms, and halo finders.

This paper is structured as follows: 
Section~\ref{sec:dbdesign} gives an overview on the database design
and describes the methods for accessing the data. In Section~\ref{sec:simu} we characterize
the current simulations in the database.
More details on how the data of these simulations are stored in the database are given in 
Section~\ref{sec:dataInDatabase}. We complement our presentation of the {\it MultiDark
Database} with a few example science cases in Section~\ref{sec:tests} and  a short summary given in
Section~\ref{sec:outlook}. Appendixes additionally provide descriptions of the employed
halofinders and the merger-tree construction.

\section{The MultiDark Database and its design}
\label{sec:dbdesign}

Databases organize large amounts of structured data for efficient
retrieval. The {\it MultiDark Database} actually is a ``relational
database''. Such relational databases organize the data in collections of
tables (originally called ``relations''), which in our case store the
different objects identified in the simulations and derived data
products.  For instance, the table containing the main
Friends-of-Friends halo catalogue (\dbtab{FOF}) consists of one record
(row) for each FOF group, with its properties mapped to the columns of
the table.

The strength of relational databases lies not only in capturing the
data itself, but in modeling possible connections between the various
datasets. Connections between the various tables are achieved by
linking rows of different tables with an unique identifier, which
points from a row  in one table to another row in the other table. Such ``foreign
keys'' establish links between the various tables in the simulation
database. For instance, the table containing the main \dbtab{FOF}
 catalogue  contains a column with a label
\dbcol{fofId} for each FOF group. This \dbcol{fofId} is used again in
the \dbtab{FOFParticles} table, which lists the simulation particles
that constitute the FOF group (see Section
\ref{sec:extractparticles} for more details).

Another important feature of relational databases is the powerful SQL
query language supported by them. As already mentioned above, SQL is
used to filter the main data products and retrieve exactly those
subsets one is interested in. SQL queries are expressed in terms of
the tables and their interrelations in the database\footnote{For an
  overview and tutorial of this language see for example
  http://www.w3schools.com/sql/default.asp.  See also the demo video
  on http://www.multidark.org/Help?page=demo/index.}.  These queries
are interpreted by the database engine and compiled into execution
plans that access the data in an optimal way to retrieve the results.
The language therefore allows users -- especially those not
intimately familiar with the data format of the simulation -- a far
more direct path from a science question to an executable expression than a
standard scripting or programming language would. I/O, looping,
optimization etc. are all handled by the database engine and the user
has no need to know about this. Moreover, the abstraction provided by
the relational model provides users with a more uniform interface even
between different databases than using file based access.

As will be explained in Sections \ref{sec:simu} and \ref{sec:dataInDatabase},
the {\it MultiDark Database} now
contains two simulations, each in a separate database with several database tables.
Users retrieve data from these tables by performing queries using
SQL.  In our case, Microsoft's SQL dialect T-SQL \citep[][]{TSQL} is
required (as is the case for the Millennium Database 
and the SDSS SkyServer).  This allows users to employ some extensions,
e.g. stored procedures to perform often used parts of queries.
Furthermore, the Spatial3D library \citep[][]{Lemson2011}, a
collection of custom data types, procedures and functions written in
C\# is provided, which simplifies and substantially speeds up queries for rectangular,
spherical, and more general volume shapes.

In order to increase the efficiency of queries on large tables,
indexes for already known query patterns were generated: for the
unique identifiers linking the tables, the mass where applicable, and
spatial coordinates. More indexes will be added for further use cases,
and for the most common query patterns of the {\it MultiDark Database} as
they emerge.

Interactive data access is provided via the MultiDark website\footnote{The
  interface was developed within the German Astrophysical Virtual
  Observatory \citep{GAVO}.}, at
\texttt{http://www.multidark.org}. Users should
register using the web form at this page, because registered users 
obtain full access to all data. However, similar to the {\it Millennium Database}, unregistered
access is enabled to a public mini-version of the MDR1-database,
featuring all tables of MDR1 for a subvolume of the MultiDark
simulation of about $(100\,h^{-1}{\rm Mpc})^3$ in the \db{miniMDR1}
database. This allows the interested user to get an overview of
capabilities of the {\it MultiDark Database} before registering. The
miniMDR1 database also serves as a test and development environment for
more elaborate queries.

The web application is designed for interactive use. SQL queries are submitted
directly via the Query Form and results are either viewed in the
browser, plotted with
VOPlot\footnote{\url{http://vo.iucaa.ernet.in/~voi/voplot.htm}} or
retrieved in various other formats (e.g. CSV-table,
VOTable\footnote{\url{http://www.ivoa.net/Documents/VOTable/}}). This
interactive interface has its limitations though, since browsers do not react
kindly to the task of e.g. rendering some megabytes of ASCII-text.
Therefore, registered users can store query results into their private
database for further use.  As already pointed out, SQL queries can
take a long time, and not well formed queries do this often.
Therefore, a limit on the query time and the private
table space is imposed.  Extending private table space or time limits for longer running queries
is possible by contacting support.

Scripted access, either for retrieving large results of queries using the
UNIX tool {\tt wget}, or for use with graphical tools like {\tt Topcat}, {\tt IDL}, or for doing
statistical analyses of radial profiles is provided by 
another servlet available at
\texttt{http://wget.multidark.org/MyDB}. Again, the documentation
provides many examples and usage hints. A web-page with often used queries
is also available.

\section{Simulation Data}
\label{sec:simu}

The Data Release 1 of the {\it MultiDark Database} contains data of two
different cosmological simulations. For each of these simulations
separate databases exist, so that further simulations and
post-processing results can be incorporated easily. 
The MultiDark Run1 and Bolshoi simulations are
complementary to the Millennium I and II simulations. All four
simulations follow the clustering of roughly the same number of dark
matter particles (8-10 billion) but within different simulation volumes, using
different cosmological parameters and different cosmological
codes. In tables \ref{t:params-cosmology} and
\ref{t:params-bolshoi-mdr1} we summarize and compare these parameters.

\begin{table}[h!t]
\begin{center}
\caption{Cosmological parameters of different simulations}
\medskip
\begin{tabular}{|l|l|l|l|}
\hline
Parameter 		& MDR1/Bolshoi& Millennium & Description\\
\hline
$h$	 		& 0.70 	& 0.73&Hubble parameter\\
$\Omega_\Lambda$	& 0.73& 0.75	& density parameter for dark energy\\
$\Omega_m$ 		& 0.27& 0.25	& density parameter for matter \\
         		&     & 	& \qquad (dark matter+baryons)\\
$\Omega_b$	 	& 0.0469 & 0.045& density parameter for baryonic matter\\
$n$			& 0.95 	& 1.0&slope of the power spectrum\\
$\sigma_8$		& 0.82 & 0.9& normalization of the power spectrum\\
\hline
\end{tabular}
\label{t:params-cosmology}
\end{center}
\end{table}

The Bolshoi simulation (from Russian ``grand, great'') has a simulation
volume of (250\,$h^{-1}$\,Mpc)$^3$ and contains $2048^3\approx 8.6\cdot 10^9$ particles
\citep{Klypin.Trujillo-Gomez.Primack.2010}. It has been performed in
2009 at the NASA Ames Research center.  The underlying cosmological
parameters are compatible with the WMAP5 and WMAP7 data, for a
discussion see \citep{Klypin.Trujillo-Gomez.Primack.2010}. These
parameters are listed in  Table
\ref{t:params-cosmology} in comparison to the cosmological parameters
used for the Millennium runs. The Bolshoi simulation
has been performed using the Adaptive Refinement Tree (ART) code
\citep[][]{Kravtsov.Klypin.Khokhlov.1997}. The code was parallelized
using MPI libraries and OpenMP directives
\citep{Gottloeber.Klypin.Springer}. The simulation is described
in detail in \citep{Klypin.Trujillo-Gomez.Primack.2010}.

The MultiDark Run1 simulation (MDR1) \citep{Prada2011} was performed in 2010 at
the NASA Ames Research center. This simulation is designed to study
galaxy clustering for the SDSS-III/BOSS survey. It contains the same
number of particles as the Bolshoi simulation but in a (1 Gpc
$h^{-1}$)$^3$ cube and takes the same cosmological parameters given
in Table ~\ref{t:params-cosmology}. Its numerical parameters are
summarized in Table~\ref{t:params-bolshoi-mdr1}.

\begin{table}[h!t]
\begin{center}
\caption{Numerical parameters of the cosmological simulations.}
\medskip
\begin{tabular}{|l|r|r|r|r|l|}
\hline
Parameter 		& MDR1 & Bolshoi & Millennium-I & Millennium-II & units \\
\hline
Box size 		& $1000$ 	&$250$ & $500$& $100$& $h^{-1}$ Mpc \\
Number of particles 	& $2048^3$ &$2048^3$&$2160^3$ & $2160^3$ & \\
Mass resolution 	& $8.721$ &$0.135$&$0.86$& $ 0.0069$ &  $10^9$ $h^{-1}$ M$_\odot$ \\ 
Force resolution 	& $7.0$ & $1.0$ &$5.0$& 1.0 &  $h^{-1}$ kpc \\
Initial redshift 	& 65 		& 80 & 127& 127 &\\
\hline
\end{tabular}
\label{t:params-bolshoi-mdr1}
\end{center}
\end{table}

\subsection{Halo catalogues}
One of the main products derived from cosmological simulations are 
halo catalogues, which are then used for further analysis. 
They contain dark matter halos (or clusters of dark matter particles)
and their intrinsic properties, like position, velocity, mass and radius. 
Several different techniques 
for finding and defining such halos were developed. Two of those halo finders
were applied to the data in the {\it MultiDark Database} and are briefly 
described in the appendixes A and B.
The {\it MultiDark Database} provides results from the BDM (Bound Density Maximum) 
halo finder (\ref{a:BDM}) which uses a spherical 3D overdensity algorithm to
identify halos and subhalos. Additionally, results from the FOF halo finder 
(\ref{a:FOF}) are provided in the database as well. The FOF halo finder
uses the relative linking length - given in terms of the mean inter-particle 
distance - to uniquely define clusters of particles.
Built on the FOF-catalogues the {\it MultiDark Database} further contains merger 
trees (see \ref{a:MERGER}), for tracing the history of halos. BDM-based merger
trees will be provided in a later data release.

In the following sections detailed descriptions of the various halo catalogues 
contained in the {\it MultiDark Database} are given.

\subsubsection{BDM catalogues in the database}
\label{sec:BDMcatsDB}
Two different BDM catalogues were produced for different definitions of the halo radius:

\begin{itemize}
\item {\bf BDMV}: the virial mass $M_{\rm vir}$~ is defined by the solution of the
top-hat model of the growth of fluctuations in an expanding Universe with a cosmological
constant. We define the virial radius $R_{\rm vir}$ of  halos as the radius
within which the mean density is the virial overdensity $\Delta_{\rm vir}(z)$ times the mean
universal matter density $\rho_{\rm m}=\Omega_{\rm m}\rho_{\rm crit}$ at that
redshift. Thus, the virial mass is given by
\begin{equation}
    M_{\rm vir} \equiv {{4 \pi} \over 3} \Delta_{\rm vir} \rho_{\rm m} R_{\rm vir}^3 \ .
\end{equation}
 For our set of cosmological parameters, at $z=0$ the virial
radius $R_{\rm vir}$~ is defined as the radius of a sphere with an overdensity of
360 of the average matter density. The overdensity limit changes with
redshift and asymptotically goes to 178 for high $z$. The overdensity $\Delta_{\rm vir}(z)$
is given by an approximation provided by \citep{BryanNorman}.

\item {\bf BDMW}: the halo radius is defined by the
overdensity limit $\Delta_{200} =200\,\rho_{\rm crit}$.  For
our set of cosmological parameters this corresponds to $740\,\rho_{\rm m}$ at
$z=0$. It approaches asymptotically the overdensity of $200 \,\rho_{\rm m}$ at 
high redshifts.  Since this density is always larger than the
virial one, the halos of the BDMW catalogue are always smaller than the
corresponding halos in the BDMV catalogue. 
\end{itemize}

Since both halo definitions are commonly used in the literature, the BDM catalogues
for both values are given.

The BDM halo catalogues provide a lot of information: each halo and subhalo is characterized with 23 parameters. The list of these
parameters is given on the website of the {\it MultiDark Database}
and is also described in  \ref{a:BDM}. In addition to coordinates, 
peculiar velocities and other halo properties, the database provides two masses: the mass of all
particles inside the virial radius and the mass of gravitationally bound
particles. For distinct halos the difference between the two masses is
typically at most 1-2 percent. The difference is much larger for
subhalos.  Note that most of the parameters of both subhalos and
distinct halos are defined by gravitationally bound particles.

\subsubsection{FOF catalogues in the database}
The nature of the FOF algorithm implies that FOF groups cannot
intersect with each other (Figure~\ref{f:fofsubhalos}), which means
that for a given linking length any particle is uniquely assigned to
just one FOF group (such a FOF group could be the particle
itself). With this property it is possible to create database tables
to establish a link between FOF groups and their particles (see
Section \ref{sec:extractparticles}). Furthermore, substructures always
lie completely within their host structure, since they are defined by
smaller linking lengths. Due to the unique mapping of a particle to a FOF
group one can determine unique progenitor-descendant-relations of FOF
groups as the basis for the construction of the merger tree.
  
For the resulting FOF groups no post-processing  has
been applied so far. In particular, no binding/unbinding procedure was
applied, i.e. a FOF group consists not necessarily of bound
particles. However, for a given FOF group all particle positions and
velocities can be extracted from the database for any post-processing.

\begin{table}
\begin{center}
  \caption{Linking lengths for the FOF catalogues provided in the
    database. The corresponding database table names are given in the
    first column, a more detailed description of these tables is
    provided in Table \ref{t:mdr1-tables}. The linking lengths are
    also stored directly in tables \dbtab{linkLength} and
    \dbtab{linkLengthScl}. The last column contains a characteristic 
    overdensity for the given linking length.}
\medskip
\begin{tabular}{|l|p{2cm}|p{2cm}|p{2cm}|}
\hline
Database table 	
& \hbox{linking length} \hbox{(in units of }\hbox{interparticle separation)} 	
& \hbox{level/sclevel in } \hbox{database table} 
& \hbox{overdensity } \\ \hline
FOF		& 0.17		 & 0 & 570. \\
FOF1		& 0.085		 & 1 & 3100. \\
FOF2		& 0.0425	 & 2 & 19000. \\
FOF3		& 0.02125	 & 3 & $1.2 \times 10^5$ \\
FOF4		& 0.010625	 & 4 & $9.8 \times 10^5$  \\
FOFc		& 0.20		 & - & 390.  \\
\hline
\multirow{6}{*}{FOFScl}		& 0.35		 & 0 & 94. \\
 	 & 0.32 	  & 1 & 120. \\
 	 & 0.29 	  & 2 & 160. \\
 	 & 0.26 	  & 3 & 210. \\
 	 & 0.23 	  & 4 & 280. \\
 	 & 0.20 	  & 5 & 390. \\
 	 & 0.17 	  & 6 & 570. \\

\hline
\end{tabular}
\label{t:linklengths}
\end{center}
\end{table}

\section{Data in the MultiDark Database}
\label{sec:dataInDatabase}

The following subsections describe the available data of the MDR1 and Bolshoi
simulation, how they are organized in the tables of the 
database, and some access examples. The names of the corresponding
tables are given in brackets at each section title. A more complete
overview of all tables and their relations can be found in
\ref{a:relations}, Figure~\ref{f:relations}.

\subsection{Halo catalogues - (\dbtab{BDMV}, \dbtab{BDMW},
           \dbtab{FOF\{1, 2, 3, 4\}}, and \dbtab{FOFc})}
\label{sec:haloCat}

Each BDM and FOF halo catalogue has a corresponding table in the
database. The BDM halo catalogues are \dbtab{BDMV} and \dbtab{BDMW}.
For the \dbtab{FOF} tables, the numbers denote the different linking
lengths used in the halo finding procedure. Furthermore, \dbtab{FOFc}
denotes the FOF halo catalogue with the commonly used linking length
of $0.2$.  A complete list of the linking lengths for the various FOF
halo catalogues is given in Table \ref{t:linklengths}.
For a given
linking length the overdensity of the FOF objects depends on the
concentration of the objects and therefore on mass and redshift
\citep{More.et.al.2011}. Moreover, it shows a large scatter. In order
to give an idea of the expected overdensity for the given linking
length, the last column in the table contains a characteristic
overdensity corresponding to the linking length given in the second
column. The halo catalogue tables
contain individual records for each dark matter halo or FOF group,
with all calculated properties given in the corresponding columns.

Additionally, spatial grid indexes (1024 cells per dimension) are
provided, together with the computed Peano-Hilbert key for each grid cell using
the Spatial3D library \citep{Lemson2011}, which enables a fast
retrieval of halos or FOF groups from a given region in space.

Another feature of the database is the quick retrieval of snapshot 
number and mass columns, or sorting of halos/FOF 
groups by their mass. This is important for e.g. calculating the mass
function of halos and 
its evolution in time, and can be done easily using database queries. 

\subsection{Halo profiles - (\dbtab{BDMVprof} and \dbtab{BDMWprof})}

For BDM halos, access to their inner structure is possible with the
BDM profiles stored in tables \dbtab{BDMVprof} and \dbtab{BDMWprof}
(\dbtab{V} and \dbtab{W} are defined as in Section \ref{sec:BDMcatsDB}).
These profiles consist of logarithmically spaced shells as a function
of the virial radius $R_{\mathrm{\rm vir}}$ of the halo. They
are available up to a radius of $2 R_{\mathrm{\rm vir}}$ and cover
halos with more than 100 particles. Each radial bin corresponds to a
row in the table and includes physical properties like e.g. local
density and circular velocity. Each property is given once for all the
particles enclosed by the shell, and once for the bound particles
only. The halo profiles tables allow the user to study density profiles,
rotation curves and other typical properties of the BDM halos.

The profile records are linked to the corresponding BDM halo from the \dbtab{BDM}-table by the 
halo's unique identifier \dbcol{bdmId}. To obtain the profile of a specific halo, one first retrieves
the halo's unique identifier \dbcol{bdmId} and then searches for all the entries in the profile table with the
corresponding \dbcol{bdmId}.

\subsection{Merger trees - (\dbtab{FOFMtree})}
\label{sec:mergertrees}

\begin{figure}[h!t]
\begin{center}
\includegraphics[width=0.7\textwidth]{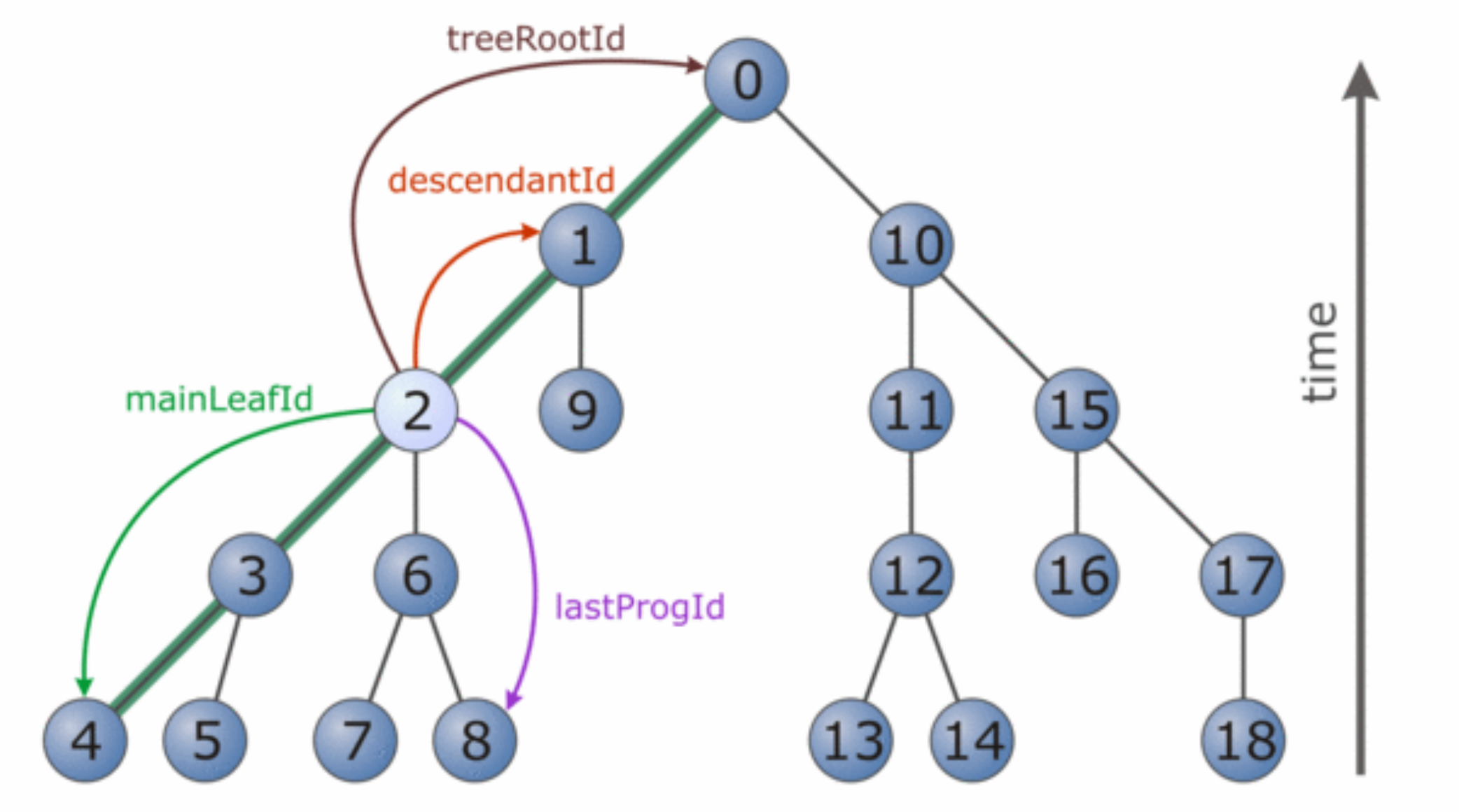}%
\caption{Merger tree: the top node (root) of the tree represents a
  halo or FOF group at redshift $z=0$. From there, branches reach
  backwards in time to its progenitors, i.e. the timeline goes
  from bottom to top. The numbers at each node indicate the
  depth-first order, with the most massive progenitors being on the
  leftmost side of each sub-tree. These form the main branch
  (e.g. the thick green line for the tree root (0)) of the corresponding node.
  The identifiers (ids) drawn here for one example node (2) are stored
  in the database table (see text for further explanations). }
\label{f:mergertree}
\end{center}
\end{figure}

Within the currently accepted ${\Lambda}CDM$ cosmology, dark matter
halos merge from small clumps to ever larger objects. This merging
history can also be traced in cosmological simulations and is stored in
the form of merger trees (see illustration, Figure
\ref{f:mergertree}). These merger trees provide a description of the
assembly history of a dark matter halo and can be used as a basis for a
semi-analytic model describing the baryonic properties of the galaxies
evolving in a halo.

Merger trees are built for a subset of halos (or FOF groups) which
exist at redshift 0 and exceed a certain mass limit. From such a
root-halo (at the top in Figure \ref{f:mergertree}), the branches go
to each of its progenitors, reaching backwards in time, with the most
massive progenitor being visited first (main branch). A typical
question is to retrieve merger trees rooted in a given halo or set of
halos. To answer such queries efficiently the same algorithm is
employed as for the Millennium database \citep{2006ADASS}. Once
the tree is built, its nodes are sorted according to a depth-first
search. This depth-first order rank is used to construct unique
identifiers and specific foreign keys, which enable quick access to
the complete merger history for each halo.

The merger-tree tables contain the following merger
tree identifiers for each FOF group:
\begin{itemize}
\item \dbcol{treeRootId}: ID of the top node in the merger tree,
  i.e. the final descendant at redshift $z=0$ (root halo), calculated based on the
  number of the halo/FOF group in the corresponding halo catalogue
  (start line number with 0):
    \begin{equation}
    	\mbox{\dbcol{treeRootId}} = (\mbox{rank in file} + 1) \cdot 10^8 
    \end{equation}
\item \dbcol{fofTreeId}: unique identifier for each FOF group, based on \dbcol{treeRootId} and rank in depth-first order:
    \begin{equation}
    \mbox{\dbcol{fofTreeId}} = \mbox{\dbcol{treeRootId}} + (\mbox{rank in merger tree}) 
    \end{equation}
    Thus, the tree membership is encoded directly in the
        \dbcol{fofTreeId} for each group.
      \item \dbcol{descendantId}: identifier (\dbcol{fofTreeId}) of
        the direct descendant of a FOF group (i.e. forward in time,
        into which the FOF group will grow/merge)

\item \dbcol{mainLeafId}: identifier (\dbcol{fofTreeId}) of the
  \emph{last} FOF group of the main branch, along the most massive
  progenitors; enables a quick retrieval of e.g. the accretion history
  of a FOF group etc. by querying for all progenitors until the FOF
  group with the \dbcol{mainLeafId} as \dbcol{fofTreeId} is
  encountered. If the halo of the top node in
  Figure~\ref{f:mergertree} is denoted as \dbcol{topHalo}, a schematic query for
  the main branch of this halo would look like:
\begin{verbatim}
select * from FOFMtree 
    where fofTreeId between topHalo.fofTreeId and topHalo.mainLeafId
\end{verbatim}
and returns the records for FOF groups no. 0, 1, 2, 3 and 4 (for the complete query, 
  consult the
  ``Very useful queries'' Nr. 5.2 on the {\it MultiDark Database} webpage).

\item \dbcol{lastProgId}: identifier (\dbcol{fofTreeId}) of the \emph{last} FOF group in the tree; queries 
for all FOF groups with \dbcol{fofTreeId} between the \dbcol{treeRootId} and the \dbcol{lastProgId} 
will return the complete merger tree.

\end{itemize}

\begin{figure}[h!t]
\begin{center}
\includegraphics[width=0.7\textwidth]{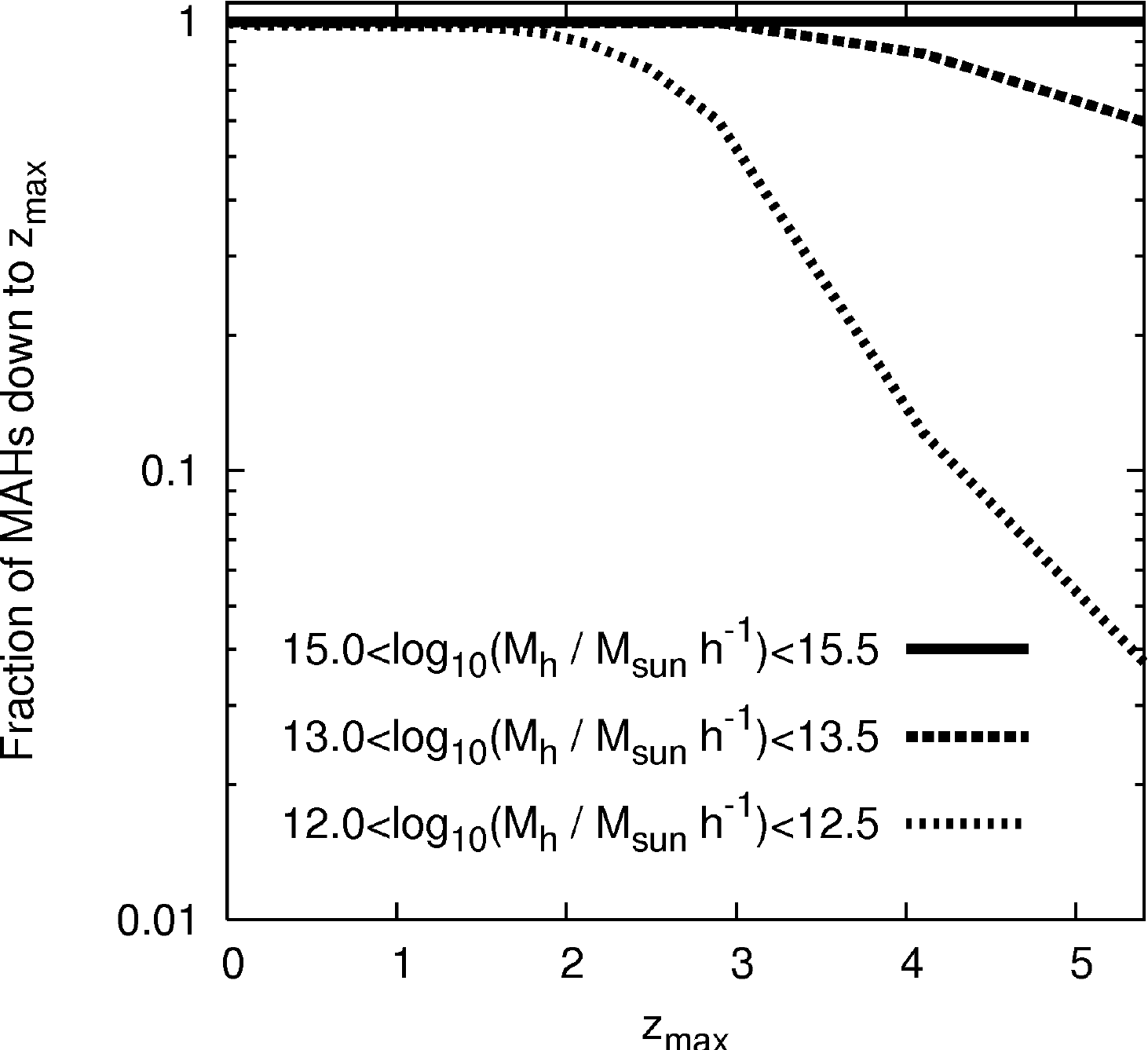}%
\caption{The fraction of FOF merger trees for the MultiDark simulation 
having a main branch followed down
to a given redshift for three different mass bins. All trees have been
constructed down to a maximum redshift of $z=5.4$.}
\label{trac-crop}
\end{center}
\end{figure}

The merger trees are determined for all halos with more than 200
particles at redshift $z=0$. They end at a certain redshift if the
main progenitor of a given halo is below the detection threshold of 20
particles. Fig. \ref{trac-crop} shows for three different mass bins of
halos at $z=0$ the fraction of halos for which the main
branch of the merger tree can be followed down to redshift $z_{\rm
max}$.

\subsection{Substructures - (\dbtab{FOFScl}, \dbtab{BDMV}, and \dbtab{BDMW})}

Small dark matter (sub)halos are embedded in larger ones, which in
turn may reside in even bigger halos. Once these multi-level subhalos
are found, their hierarchical structure can be represented by a
substructure tree, in much the same way as the build-up process of
halo formation is usually expressed with merger trees. However, this
is only possible where the information of sub-substructures is
available, as for FOF groups of different linking lengths.  For the
BDM catalogues only the \dbcol{bdmId} of the host halo for each
subhalo (in column \dbcol{hostFlag}) is available.

\begin{figure}[h!t]
\begin{center}
\includegraphics[width=0.7\textwidth]{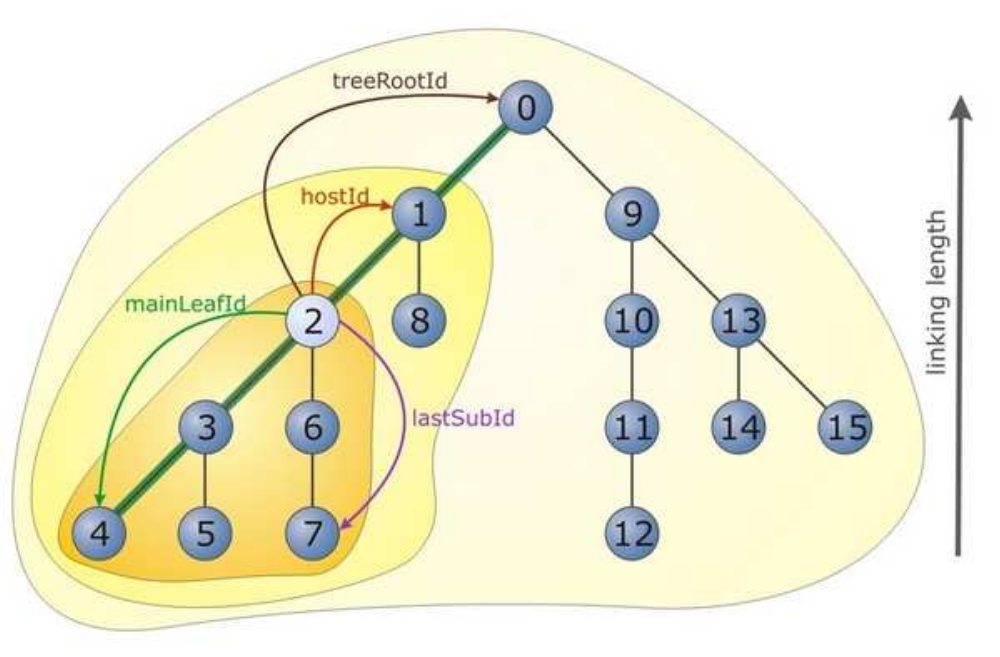}%
\caption{Substructure tree: the top node as the root of the tree
  represents the biggest object (with largest linking length). Each
  row contains FOF groups of smaller linking lengths, which are
  substructures of their host(s). Additionally FOF substructures are
  sorted by mass from left to right for each FOF group. The numbers
  indicate the ranking of FOF groups in a depth-first ordering. The
  thick green line marks the main branch of the tree root (0). The
  identifiers are constructed and used in the same way as for merger
  trees (see Section \ref{sec:mergertrees}).}
\label{f:substructuretree}
\end{center}
\end{figure}

For the FOF halo catalogues substructure trees at redshift $z=0$ are
provided.  In such a substructure tree the root node corresponds to
the biggest halo (i.e. with lowest density threshold, largest linking
length), followed by successively smaller halos in the next
substructure level. The FOF (sub)halos of the tree are sorted in a
depth-first order, with the most massive substructure as the first node
of each new level, so that the main branch can be retrieved in the
same way as for a merger tree (see Figure~\ref{f:substructuretree} and
Section~\ref{sec:mergertrees}). The necessary tree identifiers are
constructed like those for the merger trees and stored in the database tables
\dbtab{FOFSub} and \dbtab{FOFScl}. They are only renamed to fit the
substructure context (also see Figure~\ref{f:substructuretree}):
\begin{itemize}
\item \dbcol{fofTreeId} $\leftrightarrow$ \dbcol{fofSubId}
\item \dbcol{descendantId} $\leftrightarrow$ \dbcol{hostId}
\item \dbcol{lastProgId} $\leftrightarrow$ \dbcol{lastSubId}
\end{itemize}

\subsection{Simulation particles - (\dbtab{particles} and \dbtab{FOFParticles\{1, 2, 3, 4\}})}
\label{sec:extractparticles}
The {\it MultiDark Database} contains not only halo catalogues and many
related data sets like substructures and merger trees, but as one of
its main new features also the complete raw simulation data at certain
redshifts. For each of these
snapshots the full set of 8.6 billion particles is available along
with their positions and velocities\footnote{The {\it Millennium Database}
provides the particles for all 64 snapshots of the smaller ($~20$
million particles) milli-Millennium database.}.

This particle information can be used to study the particle
distribution in certain regions, e.g. in the environment of a selected
dark matter halo.  For accelerating such spatial queries the {\it
Spatial3D library} is used \citep{Lemson2011}. This library is
written in C\# and its functions and data types are available from
within T-SQL.  It employs a Peano-Hilbert space-filling curve
subdividing the box into a $1024^3$ grid.  Each particle has a column
describing the grid cell it is in. The same procedure was applied to
the halo catalogues. The use of the library is not completely
standard and example queries are provided on the web site.

For the FOF catalogues, stored in tables \dbtab{FOF} -- \dbtab{FOF4} and \dbtab{FOFScl},
additional tables contain particles of the snapshot at redshift $z=0$
linked to their corresponding FOF halos (tables \dbtab{FOFParticles}
-- \dbtab{FOFParticles4}). This allows users to easily extract particles for
a given FOF halo and e.g. calculate additional properties, or
recalculate some given quantities with alternative
methods\footnote{The
same method of linking particles to their halos has been applied
for the MMSnapshots of the Millennium database.}. The
\dbtab{FOFParticles}-tables can even be used to cross-check the
substructure information given in table \dbtab{FOFSub}: a substructure
of a FOF halo always lies completely within the host halo (since it
has smaller linking length) and thus each of its particles also
belongs to its host. By joining \dbtab{FOFParticles}-tables for
different linking lengths, one can get a list of substructures (or
hosts) for a given FOF halo, independent of the
\dbtab{FOFSub}-table\footnote{However, such queries become often quite
  expensive in terms of compute time, so it is recommended to use the
  \dbtab{FOFSub}-table for extensive substructure studies.}.

The information for the particles of a BDM halo cannot be retrieved as
directly as for the FOF halos. Since no table linking BDM halos with
their particles is currently provided, all the particles in the halo's
region (up to its virial radius) are only accessible by using the spatial
coordinates of the halo and the {\it Spatial3D library} to exctract a region
around the halo's center.

\section{Examples of using the database}
\label{sec:tests}

The {\it MultiDark Database} enables users to analyze many aspects of cosmology
and galaxy evolution. It will also help to interpret large
state-of-the-art observational data sets. The following list gives some
possible examples of analysis using data in the database:

\begin{itemize}
\item properties of halos (radial profile, concentration, shapes),
\item evolution of the number density of halos, essential for
normalization of Press-Schechter-type models,
\item evolution of the distribution and clustering of halos in real and redshift space, for
comparison with large-scale galaxy/QSO surveys,
\item accretion history of halos, assembly bias (variation of large-scale clustering with
assembly history), and correlation with halo properties including angular momentum and
shape,
\item halo statistics including the mass and velocity functions,
  angular momentum and shapes, subhalo numbers and distribution, and
  correlation with environment,
\item void statistics, including sizes and shapes and their evolution, and the orientation of
halo spins around voids.
\item quantitative descriptions of the evolving cosmic web, including applications to weak
gravitational lensing,
\item preparation of mock catalogs, essential for analysis of SDSS and
  other new survey data (SDSS-III/BOSS, DES, Planck),
\item merger trees, essential for semi-analytic modeling of the evolving galaxy population,
including models for the galaxy merger rate, the history of star formation and galaxy colors
and morphology, the evolving AGN luminosity function, stellar and AGN feedback, recycling
of gas and metals, etc.
\end{itemize}

Here we give some examples in more detail.

\subsection{Example 1: Velocity function} 
A novel feature of the {\it MultiDark Database} is the access to
the profiles of different physical parameters (density, velocity, etc)
for each of the halos found in the BDM tables.  In particular, in this
example we show how to obtain the average radial velocity profile
for halos of vastly different masses, from galaxy-size halos to
clusters.  We used many hundreds of halos for each mass range. This
simple query will allow users to study the infall of material beyond the
formal virial radius. For group- and cluster-sized halos there are
large infall velocities, whose amplitude increases with halo mass. No
infall is seen for galaxy size halos as reported by \citep{2006ApJ...645.1001P}.

Example~1 can be written using the following SQL statement:
\begin{verbatim}
    with halos as (
        select  Mvir, Rvir, bdmId
        from MDR1..BDMV
        where snapnum=85 and Mvir between 1e12 and 1.1e12
    )
    select  power(10.000000,(0.05*(0.5+floor(log10(p.R_Rvir)/0.05)))),
            avg( 
                p.Vrad /
                  ( sqrt(
                         6.67428e-8 * halos.Mvir * 1.988e33 / 
                         (halos.Rvir * 3.0856e24)
                     ) / 100000
                  ) 
            )
    from halos, MDR1..BDMVprof p
    where p.bdmId = halos.bdmId
    group by floor(log10(p.R_Rvir)/0.05) 
\end{verbatim}

Using the data retrieved with this statement, Figure~\ref{f:velplot} was generated:

\begin{figure}[h!t]
\begin{center}
\includegraphics[width=0.9\textwidth]{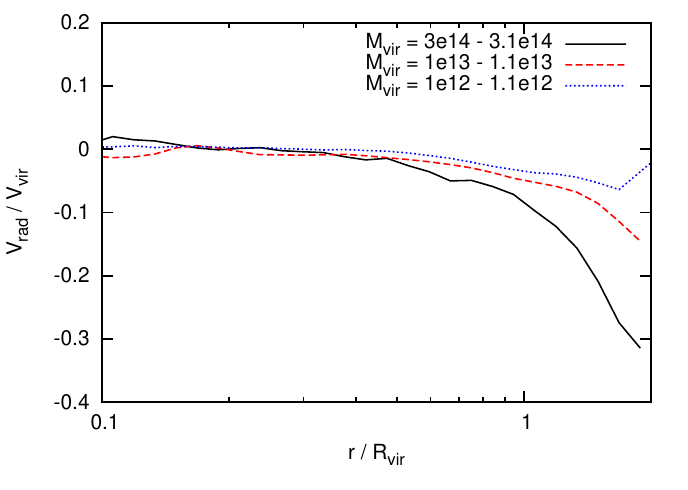}%
\caption{The plot shows the data retrieved with the statement in
  Example~1. We show average radial velocities for halos with
  different virial masses. The velocities are practically zero within
  1-2 $R_{\rm vir}$ for halos with mass $10^{12}h^{-1}M_\odot$. The situation
  is different for group- and cluster-sized halos. For these massive
  halos significant infall velocities are found. Their amplitude
  increases with halo mass.}
\label{f:velplot}
\end{center}
\end{figure}
\clearpage

\subsection{Example 2: Access to particles}
Another novel feature of the {\it MultiDark Database} is the access 
to the complete particle data of snapshots.  As an example we will retrieve
all particles which belong to the largest FOF object ({\it supercluster}) found at the
largest linking length, $ll = 0.35 $, in the {\it supercluster} table
of the MultiDark simulation. This object has a low overdensity of
about 94 (see Table \ref{t:linklengths}) and consists of 791743
particles. Its mass is $6.9\times 10^{15}h^{-1} M_{\odot}$. These
particles have been extracted from the database using the following
query:

\begin{verbatim}
    with mostMassiveCluster as (
        select top 1 * from 
                MDR1..FOFScl
            where snapnum=85 and sclevel=0
            order by mass desc
    ),
    fofClustParticles as (
        select fP.* from 
                MDR1..FOFSclParticles_85_l0 fP,
                mostMassiveCluster mC
            where fP.fofId = mC.fofSubId and fP.snapnum=85
    )
    select p.* from 
            fofClustParticles hP,
            MDR1..particles p
        where p.particleId = hP.particleId
\end{verbatim}

\begin{figure}[h!t]
\begin{center}
\includegraphics[width=0.45\textwidth]{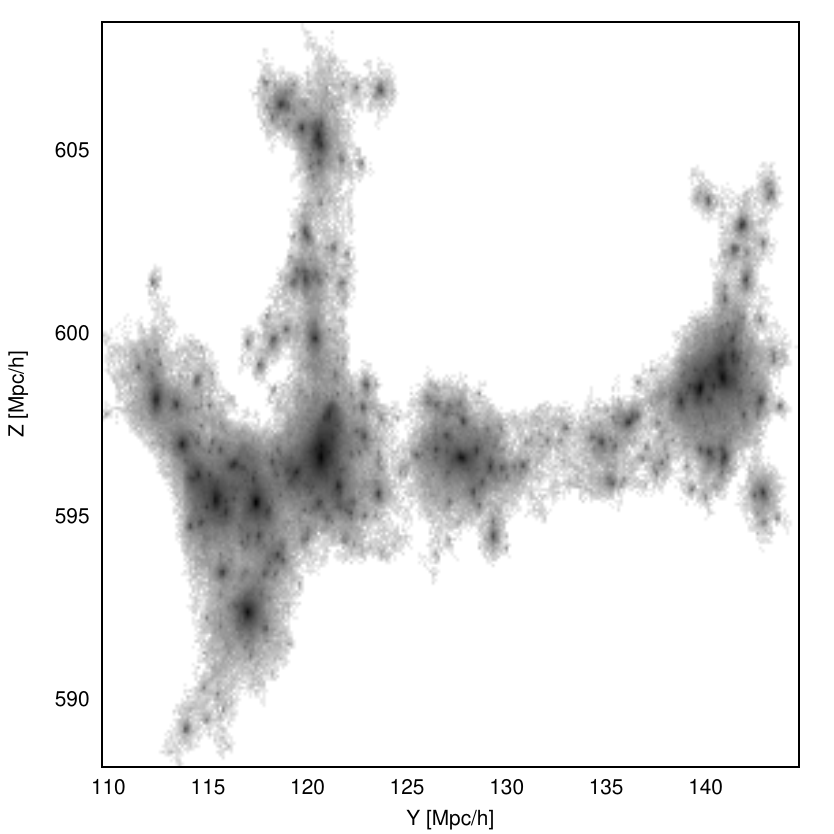}%
\includegraphics[width=0.45\textwidth]{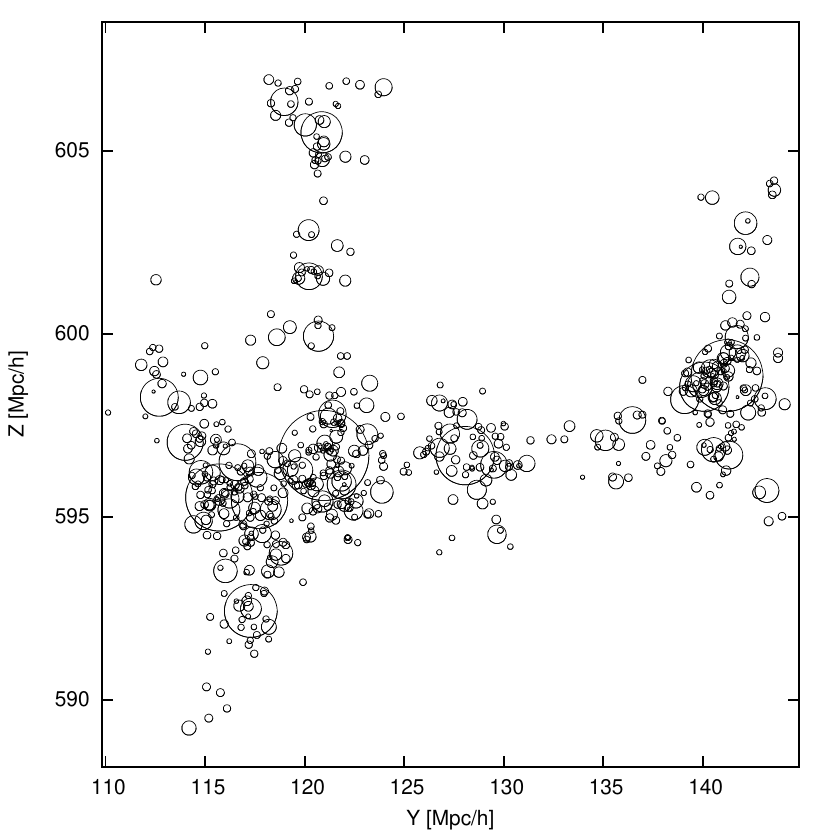}%
\hfill%
\includegraphics[width=0.45\textwidth]{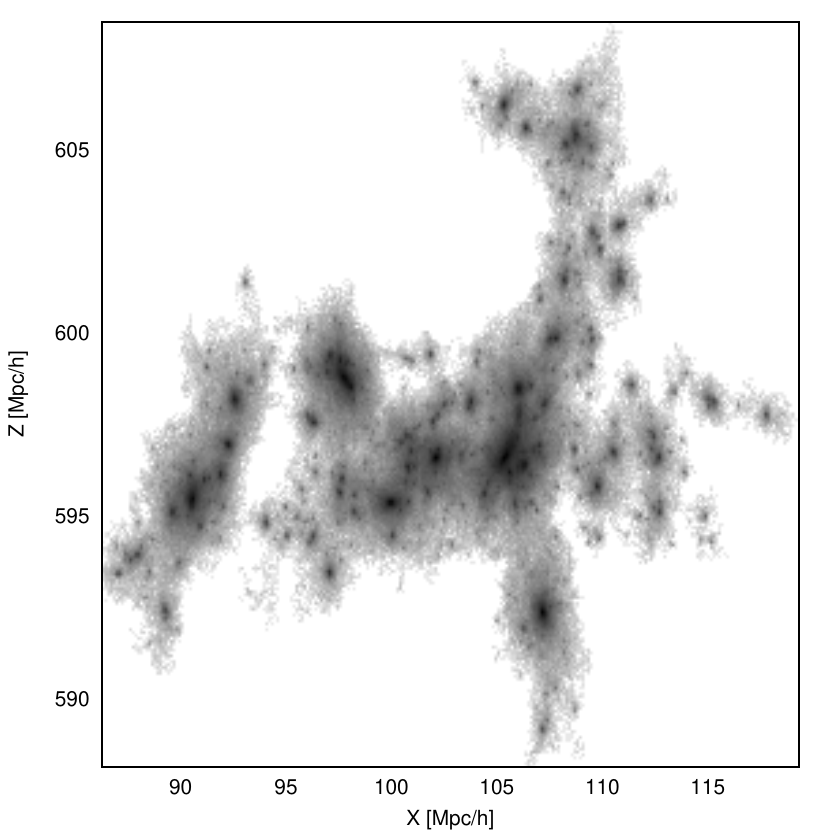}%
\includegraphics[width=0.45\textwidth]{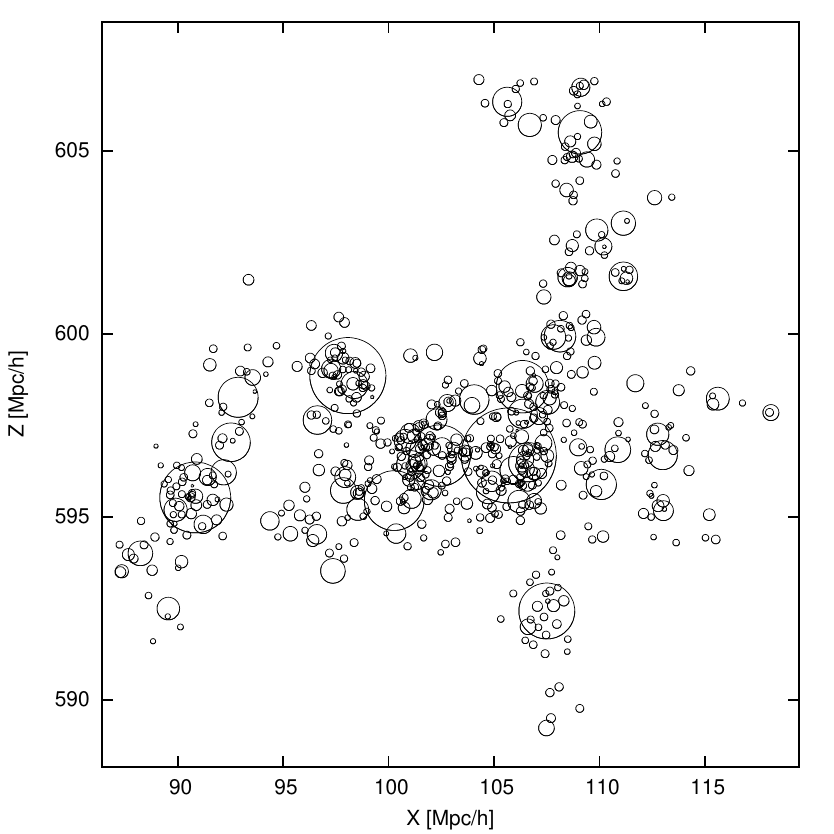}%
\hfill%
\includegraphics[width=0.45\textwidth]{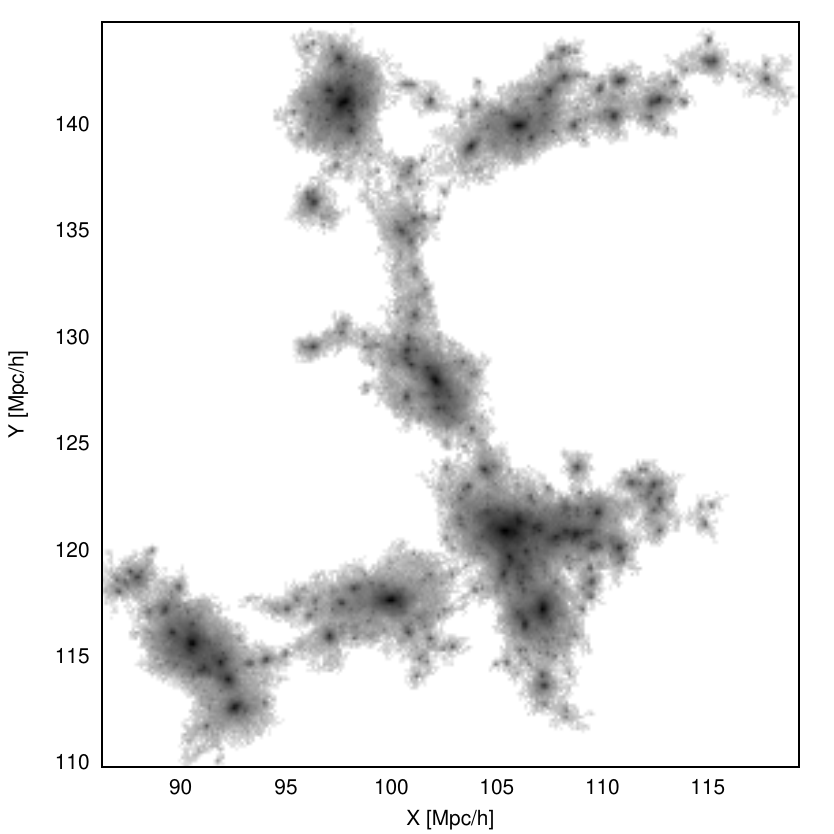}%
\includegraphics[width=0.45\textwidth]{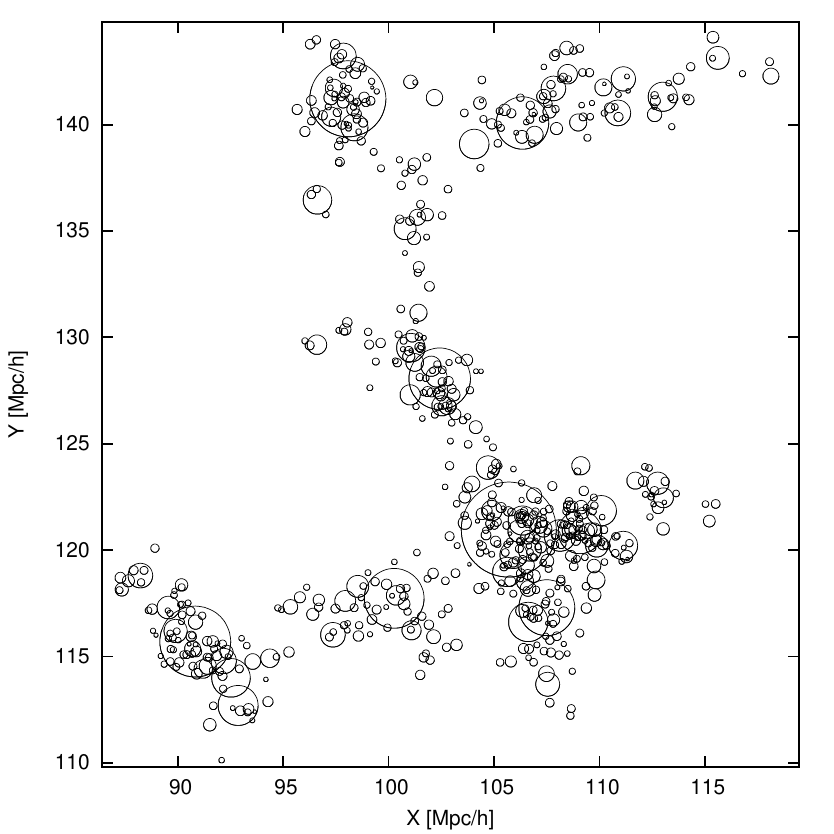}%
\hfill%
\caption{The left panes show density projections of the most massive
supercluster in the MultiDark simulation at the highest
linking length, $ll = 0.35$. The logarithm of the projected density in a
grid of cell size $130 h^{-1}$~kpc is plotted. The right panes show the objects which
the AHF halo finder identified in the same volume. The circle's radii mark
one $R_{\rm vir}$ as reported by AHF. }
\label{f:supercluster}
\end{center}
\end{figure}

The results of this query -- positions and velocities of dark matter
particles -- are retrieved by the database system in less than one
minute. Knowing the position and velocities of all these particles one
can start individual post-processing. As an example the particles of
the most massive ``supercluster'' in the MultiDark simulation are
plotted in Figure~\ref{f:supercluster} in three different projections.
In this figure, the logarithm of the
projected density in a grid of cell size $130 h^{-1}$~kpc was plotted.
The left
side of this figure shows the density distribution, whereas the right
side shows the objects which the AHF halo finder \citep{Knollmann2009}
has found in this particle distribution. The database allows the user
to download particles of one or many objects defined at a certain mean
overdensity and to run his own analysis tools, e.g. any kind of halo
finder.

\clearpage

\begin{figure}[h!t]
\begin{center}
\includegraphics[width=0.75\textwidth]{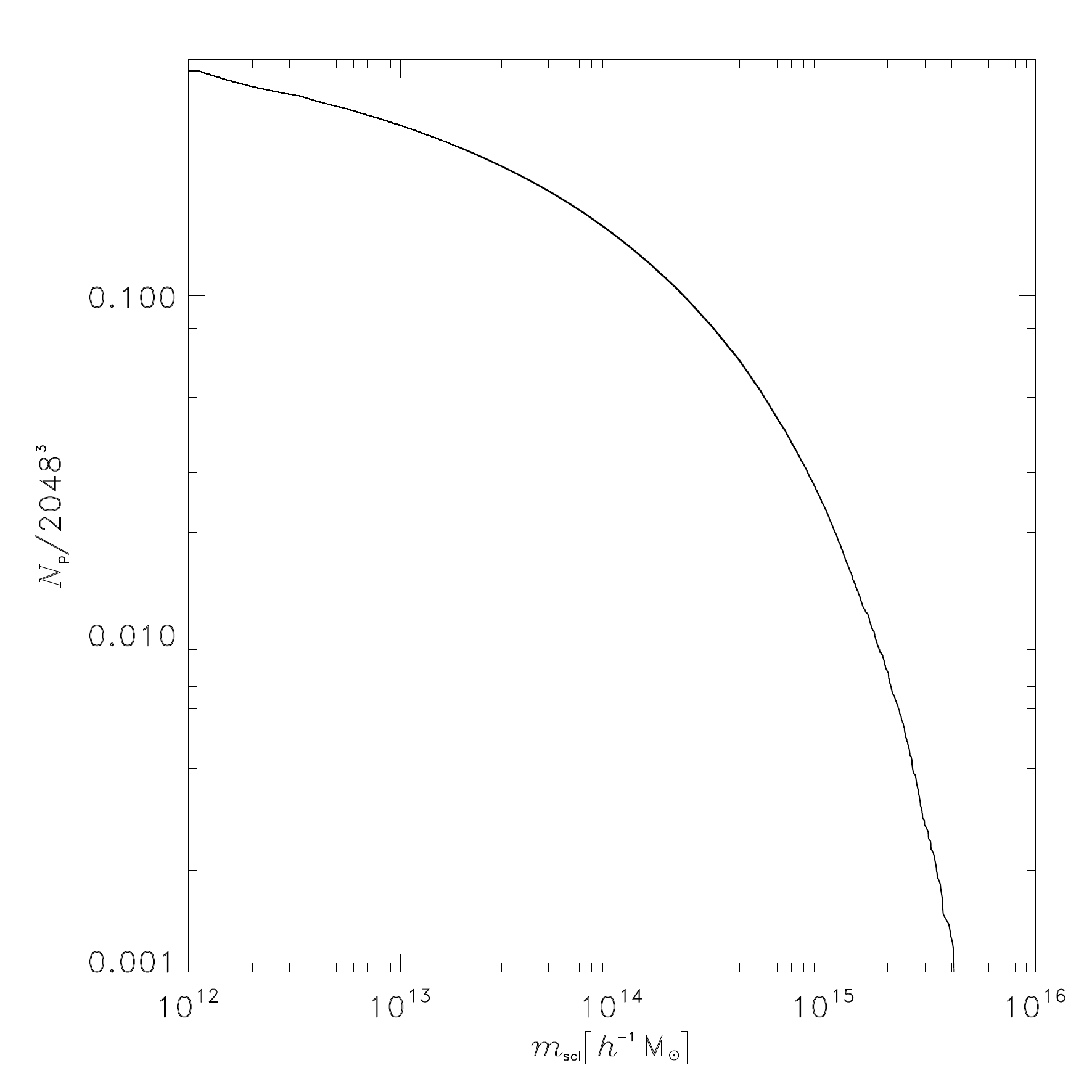}%
\caption{Cumulative fraction of particles in the MultiDark simulation
  at $z=0$ in FOF halos with masses larger than $m_{\rm scl}$. A large
  linking length $ll = 0.171 h^{-1}$ Mpc was used in this case,
  corresponding to a mean overdensity of 94.  }
\label{f:fracInHalos}
\end{center}
\end{figure}

To analyze all objects above a certain mass threshold, all the
corresponding particles from the detected ``superclusters'' can be
downloaded. As an example of such an analysis,
Figure~\ref{f:fracInHalos} shows the cumulative fraction of particles
found in FOF halos defined at overdensity 94 with masses larger than a
given mass. One can also use the downloaded particles to test or to
apply ones' own halo-finder. Since the mean overdensity is much lower
than the virial one, these FOF halos contain (for a given mass) a
complete set of spherical halos at virial overdensity. For example, to
find and analyze all spherical halos with $m_{\rm vir} > 10^{15}
h^{-1} M_{\odot}$ it would be sufficient to download all particles
from superclusters with $m_{\rm scl} > 10^{15} h^{-1} M_{\odot}$,
i.e. only about 2\,\% of all the particles (see
Figure~\ref{f:fracInHalos}).  A query which retrieves all the
particles of 1000 halos with $m_{\rm scl} > 10^{15} h^{-1} M_{\odot}$
requires 6 hours and 10 minutes. Such a query needs to be split into
individual queries for each halo due to the timeout limits.

\subsection{Example 3: Halo Mass Function}

Another very powerful feature of relational database systems is data
aggregation, such as calculating the sum of a given data set, counting
the number of data entries, or generating averages. For illustration
and to show the strength of aggregation functions, the halo mass
function is determined from the Bolshoi simulation.

Halo mass functions have been extensively studied
\citep[e.g.,][]{Jenkins2001,Tinker2008}
to obtain insight into hierarchical structure formation and the build
up of virialized objects. The mass function is also a key ingredient
in many semi-analytical models \citep[e.g.,][]{Somerville1999,Croton2006,DeLucia2007,Bower2007}.
In order to extract the halo mass function from the Bolshoi dataset at a
given redshift and for a given halo catalogue, the following SQL statement
is executed:
\begin{verbatim}
    declare @boxSizeQubed as int;
    set @boxSizeQubed = 250*250*250;
    with redZ as (    
          select snapnum 
          from Bolshoi..redshifts 
          where zred = 0.0 
    )    
    select 	0.1 * (0.5 + floor(log10(f.mass) / 0.1)) as log_mass,    
           	count(*) / 0.1 / @boxSizeQubed as num from Bolshoi..FOF f,  
           	redZ 
    where f.snapnum = redZ.snapnum 
    group by floor(log10(f.mass) / 0.1) 
    order by log_mass
\end{verbatim}
For the BDM catalogs an additional constraint selecting only distinct
haloes (i.e. \db{f.hostFlag=-1}) needs to be added to the \db{where}
clause of the statement.

\begin{figure}[h!t]
\begin{center}
\includegraphics[width=0.9\textwidth]{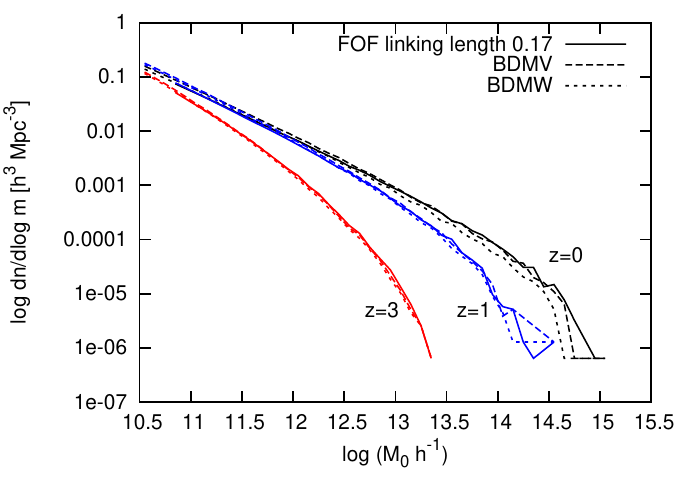}%
\caption{Halo mass function for the Bolshoi simulation derived from the FOF halo
catalog with linking lengths $0.17$ and the BDMW
catalogs. The mass function is shown at three different redshifts $z=0$ (black
lines), $z=1$ (blue lines), and $z=3$ (red lines).}
\label{f:massplot}
\end{center}
\end{figure}

With the results of this query, the halo mass function
 for the FOF and BDM halo
catalogues at three different redshifts was plotted in Figure \ref{f:massplot}.

\section{Summary}
\label{sec:outlook}

We present the {\it MultiDark Database} -- a new facility to host and
analyze large cosmological simulations. The first data release makes
 the results of two
8.6-billion particles cosmological $N$-body simulations -- Bolshoi
\citep{Klypin.Trujillo-Gomez.Primack.2010} and MultiDark Run1 \citep{Prada2011}
-- available for the astronomical community.  Data from
these simulations are organized in a relational database and are
accessible through a simple web interface. SQL can be efficiently used to pose
scientific questions, as shown in the examples. The same
technology based on an abstraction of the data in terms of tables and
relations greatly facilitates their usage and enables
comparisons. It also makes the data sets fit for dissemination by
standards developed in the International Virtual Observatory
Alliance\footnote{\url{http://www.ivoa.net}} (IVOA). In particular the
{\it Table Access
  Protocol}\footnote{http://www.ivoa.net/Documents/TAP/} (TAP) targets
the publication of, and interoperability between, datasets stored in
relational databases.

For a future data release, it is planned to include raw data for
more snapshots at different redshifts. We also plan to give access to galaxy mock catalogs
for both simulations. These mocks are based on the halo abundance
matching technique \citep[see][]{Trujillo2010}. Providing galaxy mock
catalogs to the astronomical community is essential for analyzing
large scale galaxy surveys (such as SDSS-III/BOSS, DES, Pan-Starrs),
and for planning new experiments for dark energy. Finally, we plan to
add at least the data of one more simulation in a larger volume than MultiDark Run1.

\subsection*{Acknowledgements}
The Bolshoi and MultiDark (BigBolshoi) simulations were run on the NASA's
Pleiades supercomputer at the NASA Ames Research Center. AK, JP, and SG are
grateful to the staff of the NASA Ames Research Center for helping us
with the simulations, and assisting with the analysis and visualization of the outputs. 
 We acknowledge support of NASA and NSF grants to NMSU
and UCSC for supporting this part of our research. 

We thank S. Knollmann for valuable help with the AHF halo-finder and Tam\'{a}s Budavari 
for his contribution to the Spatial 3D Library. We
acknowledge the support of the Spanish MICINN Consolider-Ingenio 2010
Programme under grant MULTIDARK CSD2009-00064.  We acknowledge the MoU
between MultiDark and AIP for the construction of the {\it MultiDark
Database}. The {\it MultiDark Database} relies on the {\it Millennium Simulation
Database} implementation and its web application for online access, which
were created by the German Astrophysical Virtual Observatory (GAVO). GAVO is
funded by the German Ministry of Education and Research (BMBF).

\newpage

\appendix
\section{Bound Density Maximum (BDM) halofinder}
\label{a:BDM}

The basic technique of the BDM halo finder is described in
\cite{Klypin.Holtzmann.1997}. The code was subject to major
improvements since 1997. It uses a spherical 3D overdensity algorithm
to identify halos and subhalos.  It starts by finding the density for
each individual particle. The density is defined using a top-hat
filter with a given number of particles $N_{\rm filter}$, which
typically is $N_{\rm filter}=20$. The code finds all density maxima,
and for each maximum it finds a sphere containing a given overdensity
mass $M_\Delta=(4\pi/3)\Delta\rho_{\rm crit}R^3_\Delta$, where
$\rho_{\rm crit}$ is the critical density of the Universe and $\Delta$
is the specified overdensity.

\begin{figure}[h!t]
\begin{center}
\includegraphics[width=0.4\textwidth]{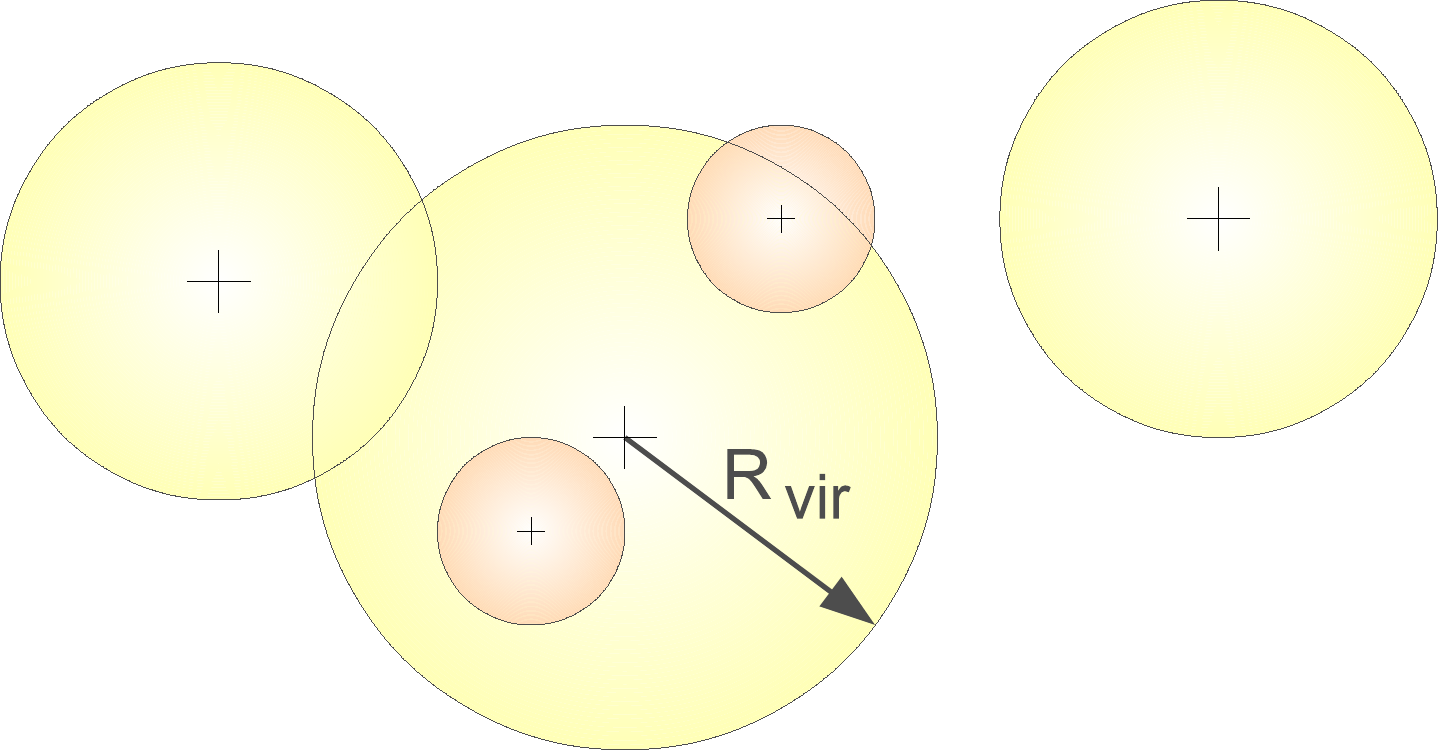}%
\caption{Relations between BDM distinct halos and subhalos. This
  example has three distinct halos (yellow) and two subhalos (light red). The right-most
  halo is a distinct halo that does not overlap with any other
  distinct halo. The left-most halo is also a distinct halo, but it
  overlaps with the largest distinct halo at the center of the picture,
  and its radius is slightly reduced. Of the two subhalos, one is
  completely inside of its ``parent'', and another subhalo is
  partially outside of its ``parent''.}
\label{f:bdmhalos}
\end{center}
\end{figure}

Among all overlapping spheres the code finds the one that has the
deepest gravitational potential. The density maximum corresponding to
this sphere is treated as the center of a distinct halo. Thus, by
construction, a center of a distinct halo cannot be inside the radius
of another one. However, peripheral regions can still partially
overlap, if the distance between centers is less than the sum of their halo
radii (see Figure~\ref{f:bdmhalos}).  Radius and mass of a distinct
halo depend on whether the halo overlaps or not with other distinct
halos. The code takes the largest halo and identifies all other
distinct halos inside a spherical shell with distances $R=(1-2)R_{\rm
  center}$ from the central large halo, where $R_{\rm center}$ is the
radius of the large halo. For each halo selected within this shell,
the code finds two radii. The first is the distance $R_{\rm big}$ to
the surface of the large halo: $R_{\rm big} =R-R_{\rm center}$. The
second is the distance $R_{\rm max}$ to the nearest density maximum in
the shell with the inner radius $\min( R_{\rm big},R_\Delta)$ and the
outer radius $\max( R_{\rm big},R_\Delta)$ from the center of the
selected halo. If there are no density maxima within that range, then
$R_{\rm max}= R_\Delta$. The radius of the selected halo is the
maximum of $R_{\rm big}$ and $R_{\rm max}$. Once all halos around the
large halo are processed, the next largest halo is taken from the list
of distinct halos and the procedure is applied again. This setup is
designed to make a smooth transition of properties of small halos when
they fall into a larger halo and become subhalos.

The bulk velocity of either a distinct halo or a subhalo is defined as the
average velocity of the 100 most bound particles of that halo or by
all particles, if the number of particles is less than 100. The number
100 is a compromise between the desire to use only the central
(sub)halo region for the bulk velocity and the noise level.

The gravitational potential is determined by first finding the mass in
spherical shells and then by integration of the mass profile. The
binning is done in $\log$ radius with a very small bin size of
$\Delta\log(R) =0.01$. An algorithm based on the pair-wise summation
was also tested. Just as one may expect, for relatively concentrated
and not-too-aspherical halos, the difference between spherical and
direct estimates are very small for the vast majority of halos (errors
are less than few percent). There are larger differences for
configurations with large and dense substructure(s). In these cases
the direct summation includes the potential energy of the substructure
in estimating the binding energy of the whole system, which is a
mistake. 
For this reason, a faster and more stable spherical estimator is
being used.

Identification of subhalos is a more complicated procedure. Centers of
subhalos can only be found among density maxima, but not all density
maxima are subhalos. The code  removes all ``fake''
subhalos: maxima of density, which do not have more than $N_{\rm
  filter}$ self-bound particles. These maxima are eliminated from the
list of subhalo candidates. An important construct for finding
subhalos are barrier points: a subhalo radius cannot be larger than
the distance to the nearest barrier point times a numerical tuning factor
called an overshoot factor $f_{\rm over}$, which is 
1.1 for the MultiDark simulation and 1.7 for the Bolshoi simulation.
The subhalo radius can be smaller than this distance. Barrier
points are centers of previously identified (sub)halos. For the first
subhalo, the barrier point is the center of the distinct halo. For the
second subhalo, it is the first barrier point and the center of the
first subhalo, and so on. The radius of a subhalo is the minimum of
 the distance to the nearest barrier point times $f_{\rm over}$ and 
the distance to its most remote bound particle.

The code starts with the density maximum and sets the first barrier
point: the center of the distinct halo. Then the bulk velocity and
profile of the gravitational potential of the subhalo are
estimated. In the next iteration unbound particles are removed and the
velocity and profiles are re-evaluated. Iterations are done until
convergence is achieved or until the number of bound particles goes
below $N_{\rm filter}$. Once a subhalo is found, a new barrier point
is added. The procedure is repeated until all subhalo candidates have been
tested.

BDM extensively uses two algorithms for rapidly finding and sorting
particles. For fast search it uses two-level link-lists. The first
level is a homogeneous mesh, which covers the whole volume, and its
size is defined by a compromise between an optimal search radius
(defined by $N_{\rm filter}$ and the overdensity limit $\Delta$) and
the available computer memory. In order to speedup the search in dense
regions, a second level of the link-list is created in regions where
the number of particles in the first link-list cell exceeds $15
N_{\rm filter}$ particles. The cell-size of the second-level mesh is 8
times smaller than the first-level one. The code also uses a {\it
  partial} ranking algorithm for finding quantities such as the most
bound particles or the particles that are closest to halo centers.

The code uses domain decomposition for MPI parallelization and OpenMP
for parallelization inside each domain.

The BDM halo catalogues provide numerous parameters for each halo and
subhalo: each halo is characterized with 23 parameters. In addition to
coordinates and peculiar velocities, the halo finder provides two
masses: the mass of all particles $M_{\rm tot}$ inside the virial radius and the mass of
gravitationally bound particles $M_{\rm vir}$. Here is a list of parameters that
require some explanations:

\begin{itemize} 

\item The offset parameter $X_{\rm off}$ is defined as the distance
  between the center of a halo and the center of halo mass $R_{\rm
    cm}$. It is given in units of the halo radius: $X_{\rm off} =
  R_{\rm cm}/R_{\rm vir}$. This parameter is often considered as a
  measure of the degree of halo relaxation.

\item The 3d rms velocity of particles $V_{\rm rms}$ relative to the halo center 
\begin{equation}
V^2_{\rm rms} = \frac{\sum_i m_iV_i^2}{\sum_i m_i}.
\label{eq:vrms}
\end{equation}
This parameter gives the kinetic energy of the halo: $E_{\rm kin}=
M_{\rm bound}V^2_{\rm rms}/2$. In combination with another parameter
provided by the code, the virial ratio 
\begin{equation}
{\rm vir}R \equiv \frac{2E_{\rm  kin}}{E_{\rm pot}}-1,
\end{equation}
 one can obtain the potential energy $E_{\rm  pot}$.

\item The maximum circular velocity $V_{\rm circ}$ is defined using
  the distribution of mass $M(<R)$ inside radius $R$. The code bins all bound particles
  using very narrow spherical shells. The binning is done in constant
  increments of the logarithm of the radius with $\Delta\log R =0.01$. This yields
  a maximum relative error in the radius of about 0.02 and even a smaller error
  in $V_{\rm circ}$.  Then, the code searches for the maximum of
  circular velocity $\sqrt{GM(<R)/R}$ starting from the first bin
  containing at least 5 particles.

\item The halo concentration $C$ is defined by the halo mass $M_{\rm
    vir}$ and the maximum circular velocity $V_{\rm circ}$. The
  algorithm of \citep{Prada2011} was used to find the
  concentration. The concentration is found by numerically solving
  the algebraic equation
\begin{equation}
\left(\frac{V_{\rm circ}}{V_{\rm vir}}\right)^2 = \frac{0.2162C}{F(C)},
\end{equation}
where $V^2_{\rm vir}=GM_{\rm vir}/R_{\rm vir}$ and $F(C) =\ln(1+C)/C-1/(1+C)$.

\item The spin parameter $\lambda$ is defined here as
\begin{equation}
\lambda \equiv\frac{J E_{\rm kin}^{1/2}}{GM_{\rm vir}^{5/2}} = \frac{jV_{\rm rms}}{\sqrt{2}GM_{\rm vir}},
\label{eq:spin}
\end{equation}
where $J$ and $j$ are respectively the total and speciﬁc angular momenta of the bound halo particles 
relative to the halo center. Note that the kinetic, not the total
energy is used to define the spin parameter. If the
total energy is wanted, it can be obtained from $V_{\rm rms}$ and $R_{\rm
  vir}$.

\item The rms radius $R_{\rm rms}$ of bound particles:
\begin{equation}
R^2_{\rm rms} = \frac{\sum_i m_iR_i^2}{\sum_i m_i}.
\label{eq:rrms}
\end{equation}

\item The axis ratios and the direction of
the major axis of the halo's triaxial shape.  This information is
obtained from diagonalization of the modified tensor of inertia ${\cal
  T}_{jk}$ for all bound particles inside the halo radius:
\begin{equation}
{\cal T}_{jk} = \sum_i \frac{x_{ij} x_{ik}}{R_{i}^2},
\label{eq:inertiatensor}
\end{equation}
where $i$ is the particle index and $j$, $k = 1,2,3$. Here $x$ stands
for the position and $r$ for the distance of a particles with respect
to the halo's center \citep[see also e.g. ][ equation
(5)]{Allgood.et.al.2006}.  The code does not use any corrections of
the axial ratios to compensate for the fact that $T_{jk}$ is estimated
using a spherical region \citep{Allgood.et.al.2006}.  A correction
factor for the axial ratios should be applied. However, the correction
depends on the halo concentration: it is smaller for more concentrated
halos. If $(c/a)$ and $(b/a)$ are the small-to-large and medium-to-large
axial ratios provided by the diagonalization of the modified inertia
tensor, then the following corrections give the true axial ratios for halos
with a flattened NFW profile:

\begin{eqnarray}
\left(\frac{c}{a}\right)_{\rm cor}&=& \left(\frac{c}{a}\right)^s, \quad s = 1 + 2\max(q-0.4,0) +[5.5\max(q-0.4,0)]^3,\\
\left(\frac{b}{a}\right)_{\rm cor}&=& \left(\frac{b}{a}\right)^p,\quad p  = 1 + 2\max(q-0.4,0) +[5.7\max(q-0.4,0)]^3\\
q  &=& \frac{R_{\rm rms}}{R_{\rm vir}}, 
\end{eqnarray}

\end{itemize}

\section{Friends-of-Friends halofinder (FOF)}
\label{a:FOF}
The Friends-of-Friends (FOF) method dates back to
\citet{Davis.et.al.1985}. This method is one of the most popular
algorithms used to find objects in cosmological simulations.  The
great advantage of this method is its simplicity: The algorithm is
based on only one free parameter - the relative linking length $ll$ - which
is given in terms of the mean inter-particle distance. For a given
linking length the FOF algorithm uniquely defines clusters of
particles that contain all particles separated by distances smaller
than the linking length. In the limit of large number of particles the
boundary of a cluster of particles is given by a certain isodensity
surface. When the FOF algorithm was introduced into numerical
cosmology, the commonly used value of the linking length was $ll=0.2$,
assuming that this value corresponds to a surface overdensity of
$\approx 60$, which in turn corresponds to an enclosed overdensity of 180
in an isothermal density profile, as desired for virialized
halos in the
standard cosmology model at that time ($\Omega_m = 1$, $\Lambda =
0.$).  Since the modern ${\Lambda}CDM$ model requires at $z=0$ a higher
overdensity of about 360 with respect to the mean density ($\Omega_m =
0.3$, $\Lambda = 0.7$), and since overdensities scale with (linking
length)$^{-3}$, for these models a linking length of 0.17 is required
at $z=0$. However, the virial overdensity in these models changes
with redshift as predicted by the spherical top-hat model
\citep{1991MNRAS.251..128L}. It reaches again the value 180 of the
Einstein-deSitter universe at high redshifts (namely when the
cosmological constant is dynamically not important), thus a redshift
dependent linking length would be required. Since this contradicts the
idea of having only one parameter, most FOF halofinders use a fixed
linking length for all redshifts. Moreover, it is well known for a
long time that the overdensity of FOF objects defined with a certain
linking length has a large scatter, and on average it is larger than
expected. In fact, recently \citep{More.et.al.2011} have shown that for
a linking length of $ll=0.2$ of the mean interparticle distance the surface
overdensity of FOF groups is equal to 81.62 times the mean density in
the simulation box. Consequently, the enclosed overdensity is larger
than 180.  It also depends on the concentration of the objects
and therefore on mass and redshift. For a linking length of $ll=0.2$ it
typically scatters between 250 and 600. For a detailed discussion of
the relation between the overdensity with respect to the mean density
and the linking length, see
\cite{More.et.al.2011}. Table~\ref{t:linklengths} provides a
characteristic overdensity for all linking lengths available in the
database.

\begin{figure}[h!t]
\begin{center}
\includegraphics[width=0.8\textwidth]{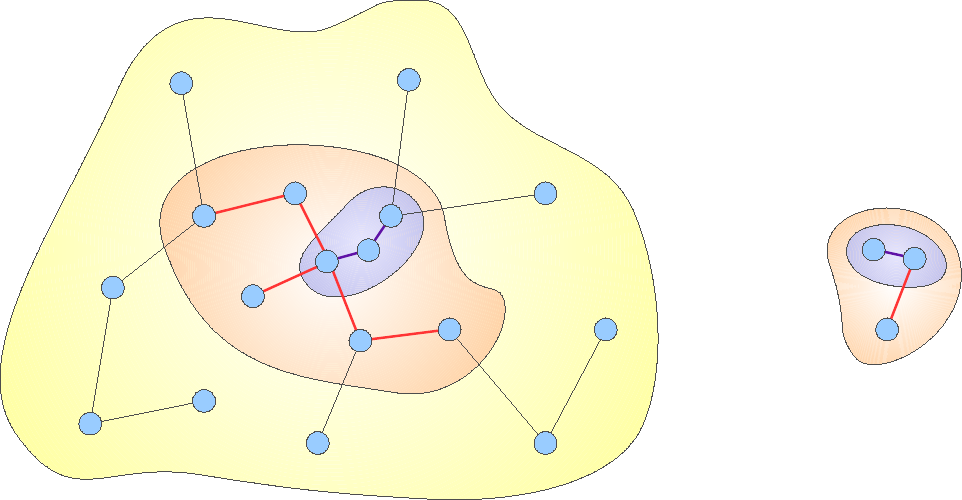}%
\caption{FOF halos constructed with larger linking length contain
  halos constructed with smaller linking length.}
\label{f:fofsubhalos}
\end{center}
\end{figure}

In order to analyze the MultiDark and Bolshoi simulations with
$2048^3$ particles a new parallel version of the hierarchical
friends-of-friends algorithm with low memory requirements was
developed. The dark matter particles in the simulation are
considered as an undirected graph with positive weights, namely the
lengths of the segments of this graph. For simplicity, it is assumed
that all weights are different. Then one can show that a unique
Minimum Spanning Tree (MST) of the point distribution exists, namely
the shortest graph which connects all points. If subgraphs cover the
graph, then the MST of the graph belongs to the union of MSTs of the
subgraphs. Thus subgraphs can be constructed in parallel. Moreover,
the geometrical features of the FOF objects, namely the fact that they
occupy mainly almost non-overlapping volumes, allow for the construction
of fast parallel algorithms. In a second step the particles are sorted
in a one-dimensional array so that each particle-cluster at any
linking length is a segment of this array. This representation yields
fast determination of FOF clusters at any linking length as well as
the substructures of FOF clusters defined at any shorter linking
length. Moreover, it is possible to determine a very fast calculation of the
progenitor-descendant-relationships of FOF objects. Since at all
redshifts FOF catalogues for several different linking lengths are
available, one can query for the corresponding substructures. In
addition, at redshift $z=0$ large linking lengths have been used to
find superclusters of FOF particles. These structure elements have
been determined with a set of linking lengths representing mean
overdensities down to 94 (see Table \ref{t:linklengths}). For all
these superclusters one has access to the raw particle data.

When the particles belonging to a given particle cluster have
been determined, different properties of this FOF group can be directly
calculated.  The database provides the center of mass of the FOF
group, its velocity, the number of particles belonging to the FOF
group, the total mass, and the velocity dispersion inside the FOF halo 
(eq.~(\ref{eq:vrms})). Since the FOF concept is by construction
aspherical, a circular velocity (as used to characterize objects found
with spherical overdensity algorithms) cannot be determined here. The
database provides two estimates of a "radius": (1) the coordinate
dispersion (rms radius) of particles $R_{\rm rms}$ defined in the same
way as in eq.~(\ref{eq:rrms}), but in this case using all particles
found by the FOF algorithm and (2) the radius of the sphere that has
the same volume as the FOF group. This volume of the particle cluster
is calculated on a grid. To save space, this volume is not included in
the database.  A mean overdensity of the FOF is provided, $\delta =
\rho_{\rm FOF}/\rho_{\rm mean} - 1$. The database also gives the
vector $J$ of the angular momentum of each FOF group. With this
vector and the total kinetic energy $E_{\rm kin}$ of the motion of 
particles relative to the halo center, the spin parameter is calculated using 
\begin{equation}
\lambda =\frac{J E_{\rm kin}^{1/2}}{GM^{5/2}}
\label{eq:spin2}
\end{equation}

Finally, the axial ratios of a FOF halo are defined as the ratios of
the main axes of the tensor of inertia of FOF particles
(eq. \ref{eq:inertiatensor}). The orientation of the halo is
characterized by three unit vectors pointing along these three main
axes.

\section{Merger trees for FOF catalogues}
\label{a:MERGER}
After finding the FOF halos in the simulation, for all the available
snapshots the merging trees are determined for all halos with more
than 200 particles at $z=0$. The construction of the trees is based on
the comparison of two consecutive snapshots. Starting at $z=0$ for
every FOF group in the catalog, $G_{0}$, all the FOF groups in the
previous snapshots are identified that share at least 13 particle
with $G_{0}$ and labeled as tentative progenitors. Then, for each
tentative progenitor, all the descendants sharing at least 13
particles are determined. Only the tentative progenitors that have the
group $G_{0}$ as a main descendant are labeled as confirmed
progenitors at that level. This procedure is iterated for each
confirmed progenitor, until the last available snapshot at high
redshift is reached. By construction, each halo in the tree can have
only one descendant but many progenitors.

It is also important to note, that the correspondence between FOF
halos and trees is not always one to one, neither are all halos
included in the trees. A possible reason is that FOF halos temporarily
disappear, because of the detection threshold of 20 particles during a
single snapshot. In this case, the branch is cut at the first snapshot
where the halo disappears. Another possible reason is that FOF halos
are sometimes linked by temporary particle bridges. In this case the
algorithm detects a halo splitting when the bridge disappears, cutting
the merger history of the less massive FOF clump at that time, while
stitching the rest of its formation history to a tree branch of the
most massive clump.

\newpage

\thispagestyle{empty}
\begin{table}
\section{Database tables - Overview}

Table~\ref{t:mdr1-tables} provides an overview of tables in databases for Multidark and Bolshoi. For a complete 
overview of the {\it MultiDark Database} tables see \url{http://www.multidark.org/MultiDark/pages/Status.jsp}.

\begin{center}
\caption{Names and description of tables in the MDR1, miniMDR1, and Bolshoi databases. }
\medskip
\begin{tabular}{|p{2.4cm}|p{4cm}|p{8cm}|}
\hline
Database Table 	& Short description & Description\\ \hline 

BDMV 		& BDM halos, $360 \cdot \rho_{back}$ 
    		& Halo catalogue using the Bound Density Maximum algorithm, for all available snapshots, 
    		calculated using the standard overdensity criterion with $360 \cdot \rho_{back}$ (background density)\\

BDMW 		& BDM halos, $200 \cdot \rho_{crit}$ 
    		& for all available snapshots, calculated using $200 \cdot \rho_{crit}$ (critical density) for defining the halo boundary\\

BDMVprof, BDMWprof 		& profiles for BDM halos
    		& corresponding halo profiles for halos from tables BDMV and BDMW\\

FOF 		& \hbox{\strut FOF groups,} \hbox{linking length 0.17}
    		& Groups of galaxy cluster size, determined using the Friends-of-Friends analysis, for all available
    		snapshots, level 0 (relative linking length 0.17)\\

FOF1 -- FOF4 	& FOF groups, smaller linking lengths (substructures)
    		& Friends-of-Friends catalogues for all available snapshots, same as FOF-table, but for smaller 
    		(relative) linking lengths, levels 1 (linking length 0.085) to 4 (linking length 0.010625). 
    		Thus these tables contain substructures of the FOF groups.\\

FOFc 		& FOF groups, commonly used linking length 0.2
    		& Friends-of-Friends catalogue for all available snapshots, computed with the commonly used linking length 0.2\\

FOFParticles 	& FOF groups $\leftrightarrow$ particles
    		& Table for connecting FOF groups from the FOF table with its particles, at the moment for redshift 0 (snapnum=85) only\\

FOFParticles1 -- FOFParticles4 	
    	& \hbox{FOF groups $\leftrightarrow$ particles,} \hbox{for FOF1 -- FOF4}
    		& Corresponding tables for connecting FOF groups from tables FOF1 -- FOF4 to their particles for redshift 0\\

FOFMtree		& merger trees for FOF
    		 & Contains identifiers to extract merger trees for galaxy clusters from FOF table\\

FOFSub		& substructures for FOF -- FOF4 
    		 & Substructure tree identifiers for building a substructure tree with FOF groups from FOF -- FOF4\\
    	
FOFScl 		& FOF superclusters
    		& Contains Friends-of-Friends catalogues for redshift 0, for 7 different (relative) linking
    		lengths between 0.35 (sclevel 0) and 0.17 (sclevel 6), i.e. much larger objects (superclusters).
    		It also contains identifiers for building substructure trees\\

particles 	& all particles, snapshot at $z=0$ for both MDR1 and Bolshoi, snapshots at $z=2.89, 1.0, 0.53$ for MDR1
    		& All simulation particles with their positions and velocities\\
\hline

linkLength 	& \hbox{linking lengths,} \hbox{FOF -- FOF4}
    		& Overview on levels and corresponding linking lengths for FOF -- FOF4 catalogues\\

linkLengthScl 	& \hbox{linking lengths,} \hbox{FOF superclusters}
    		& Overview on sclevels for superclusters and corresponding linking lengths (for FOFScl table)\\

redshifts 	& snapshots and redshifts
    		& Overview on available snapshots (snapnum) and corresponding redshifts\\
\hline

\end{tabular}
\label{t:mdr1-tables}
\end{center}
\end{table}
\clearpage

\newpage 
\begin{sidewaysfigure}
\section{Diagram of the table relations in {\it MultiDark Database}}
\label{a:relations}
\begin{center}
\includegraphics[width=0.9\textwidth]{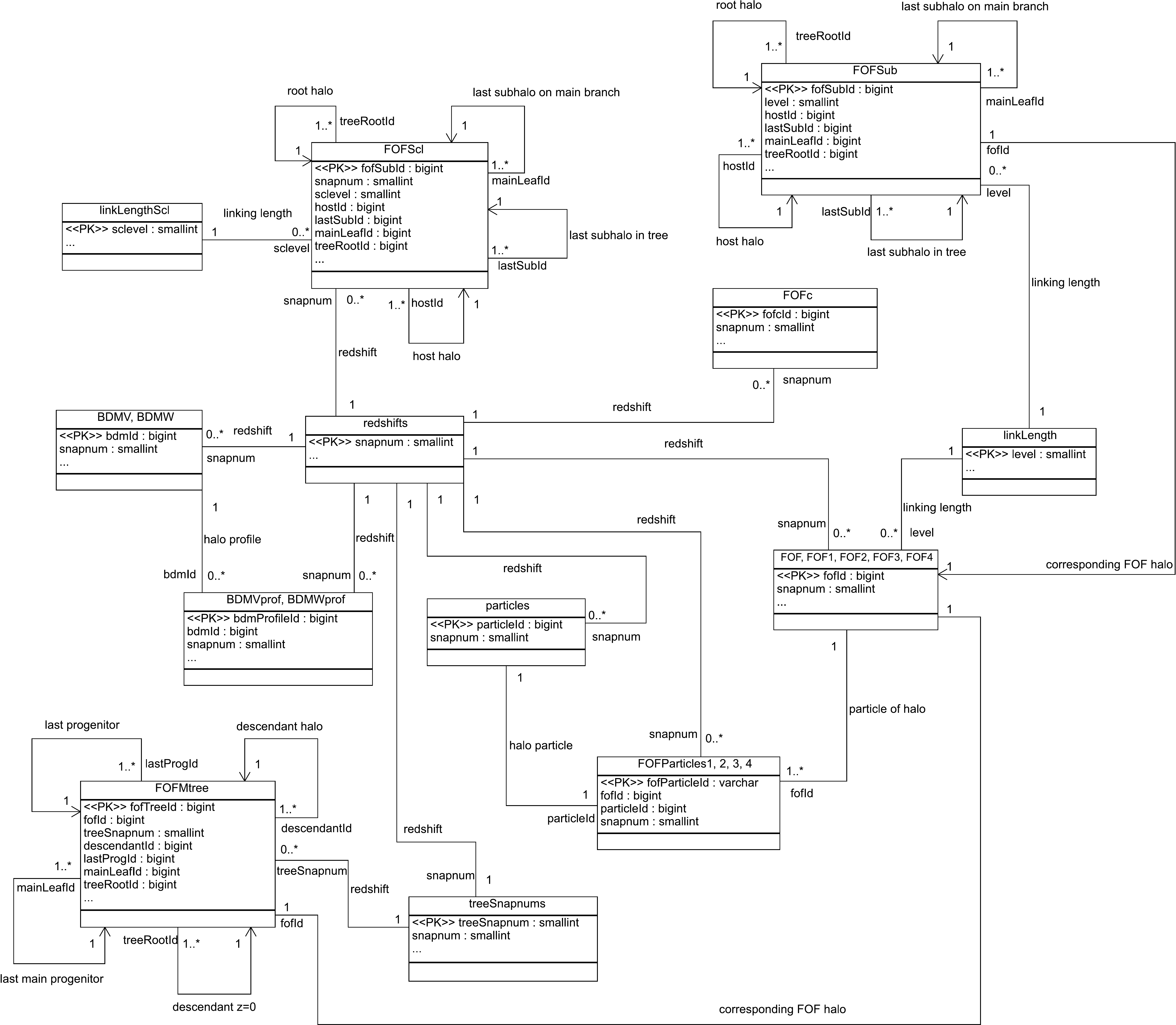}%
\caption{Relational diagram of the available tables in the MultiDark
  database.  The connecting lines mark relations between two
  tables. The connecting keys are stated along the line, together with
  the purpose of the relation. The numbers along the connecting lines
  indicate the type of the relation, for instance whether a relation
  is a one to one relation, one to many relation, or a many to many
  relation. For simplicity, only the relevant keys in the table
  objects are shown. For a complete UML diagram of the whole database
  structure, see the MultiDark webpage.}
\label{f:relations}
\end{center}
\end{sidewaysfigure}

\clearpage
\newpage

\bibliographystyle{elsarticle-harv}
\bibliography{multidarkrefs}

\begin{thebibliography}{63}
\expandafter\ifx\csname natexlab\endcsname\relax\def\natexlab#1{#1}\fi
\expandafter\ifx\csname url\endcsname\relax
  \def\url#1{\texttt{#1}}\fi
\expandafter\ifx\csname urlprefix\endcsname\relax\def\urlprefix{URL }\fi

\bibitem[{{Aarseth}(1966)}]{Aarseth1966}
{Aarseth}, S.~J., 1966. {Dynamical evolution of clusters of galaxies, II}.
  \mnras 132, 35.

\bibitem[{{Aarseth}(1969)}]{Aarseth1969}
{Aarseth}, S.~J., 1969. {Dynamical evolution of clusters of galaxies-III.}
  \mnras 144, 537.

\bibitem[{{Allgood} et~al.(2006){Allgood}, {Flores}, {Primack}, {Kravtsov},
  {Wechsler}, {Faltenbacher}, and {Bullock}}]{Allgood.et.al.2006}
{Allgood}, B., {Flores}, R.~A., {Primack}, J.~R., {Kravtsov}, A.~V.,
  {Wechsler}, R.~H., {Faltenbacher}, A., {Bullock}, J.~S., Apr. 2006. {The
  shape of dark matter haloes: dependence on mass, redshift, radius and
  formation}. \mnras 367, 1781--1796.

\bibitem[{Bower et~al.(2007)Bower, Benson, Malbon, Helly, Frenk, Baugh, Cole,
  and Lacey}]{Bower2007}
Bower, R.~G., Benson, A.~J., Malbon, R., Helly, J.~C., Frenk, C.~S., Baugh,
  C.~M., Cole, S., Lacey, C.~G., 2007. {Breaking the hierarchy of galaxy
  formation}. MNRAS 370, 645--655.

\bibitem[{{Boylan-Kolchin} et~al.(2009){Boylan-Kolchin}, {Springel}, {White},
  {Jenkins}, and {Lemson}}]{2009MNRAS.398.1150B}
{Boylan-Kolchin}, M., {Springel}, V., {White}, S.~D.~M., {Jenkins}, A.,
  {Lemson}, G., Sep. 2009. {Resolving cosmic structure formation with the
  Millennium-II Simulation}. \mnras 398, 1150--1164.

\bibitem[{{Bryan} and {Norman}(1998)}]{BryanNorman}
{Bryan}, G.~L., {Norman}, M.~L., Sep. 1998. \apj 495, 80.

\bibitem[{{Bullock} et~al.(2001){Bullock}, {Kolatt}, {Sigad}, {Somerville},
  {Kravtsov}, {Klypin}, {Primack}, and {Dekel}}]{Bullock2001}
{Bullock}, J.~S., {Kolatt}, T.~S., {Sigad}, Y., {Somerville}, R.~S.,
  {Kravtsov}, A.~V., {Klypin}, A.~A., {Primack}, J.~R., {Dekel}, A., Mar. 2001.
  {Profiles of dark haloes: evolution, scatter and environment}. \mnras 321,
  559--575.

\bibitem[{{Conroy} et~al.(2006){Conroy}, {Wechsler}, and
  {Kravtsov}}]{Conroy2006}
{Conroy}, C., {Wechsler}, R.~H., {Kravtsov}, A.~V., Aug. 2006. {Modeling
  Luminosity-dependent Galaxy Clustering through Cosmic Time}. \apj 647,
  201--214.

\bibitem[{{Croton} et~al.(2006){Croton}, {Springel}, {White}, {De Lucia},
  {Frenk}, {Gao}, {Jenkins}, {Kauffmann}, {Navarro}, and
  {Yoshida}}]{Croton2006}
{Croton}, D.~J., {Springel}, V., {White}, S.~D.~M., {De Lucia}, G., {Frenk},
  C.~S., {Gao}, L., {Jenkins}, A., {Kauffmann}, G., {Navarro}, J.~F.,
  {Yoshida}, N., Jan. 2006. {The many lives of active galactic nuclei: cooling
  flows, black holes and the luminosities and colours of galaxies}. \mnras 365,
  11--28.

\bibitem[{{Davis} et~al.(1985){Davis}, {Efstathiou}, {Frenk}, and
  {White}}]{Davis.et.al.1985}
{Davis}, M., {Efstathiou}, G., {Frenk}, C.~S., {White}, S.~D.~M., May 1985.
  {The evolution of large-scale structure in a universe dominated by cold dark
  matter}. \apj 292, 371--394.

\bibitem[{{De Lucia} and {Blaizot}(2007)}]{DeLucia2007}
{De Lucia}, G., {Blaizot}, J., Feb. 2007. {The hierarchical formation of the
  brightest cluster galaxies}. \mnras 375, 2--14.

\bibitem[{{Dubinski} and {Carlberg}(1991)}]{Dubinski1991}
{Dubinski}, J., {Carlberg}, R.~G., Sep. 1991. {The structure of cold dark
  matter halos}. \apj 378, 496--503.

\bibitem[{{Efstathiou} and {Jones}(1979)}]{Efstathiou1979}
{Efstathiou}, G., {Jones}, B.~J.~T., Jan. 1979. {The rotation of galaxies -
  Numerical investigations of the tidal torque theory}. \mnras 186, 133.

\bibitem[{{Gao} et~al.(2005){Gao}, {Springel}, and {White}}]{Gao2005}
{Gao}, L., {Springel}, V., {White}, S.~D.~M., Oct. 2005. {The age dependence of
  halo clustering}. \mnras 363, L66--L70.

\bibitem[{GAVO(2008)}]{GAVO}
GAVO, 2008. \url{http://www.g-vo.org/www/}.

\bibitem[{{Gott} et~al.(1979){Gott}, {Turner}, and {Aarseth}}]{Gott1979}
{Gott}, III, J.~R., {Turner}, E.~L., {Aarseth}, S.~J., Nov. 1979. {N-body
  simulations of galaxy clustering. III - The covariance function}. \apj 234,
  13.

\bibitem[{{Gottl{\"o}ber} and {Klypin}(2008)}]{Gottloeber.Klypin.Springer}
{Gottl{\"o}ber}, S., {Klypin}, A., 2008. {The ART of Cosmological Simulations}
  {High Performance Computing in Science and Engineering, Garching/Munich 2007,
  Transactions of the Third Joint HLRB and KONWIHR Status and Result Workshop,
  Eds.: Wagner, S.; Steinmetz, M.; Bode, A.;" Brehm, M., Springer-Verlag},
  29--+.

\bibitem[{{Iliev} et~al.(2011){Iliev}, {Mellema}, {Shapiro}, {Pen}, {Mao},
  {Koda}, and {Ahn}}]{Iliev2011}
{Iliev}, I.~T., {Mellema}, G., {Shapiro}, P.~R., {Pen}, U.-L., {Mao}, Y.,
  {Koda}, J., {Ahn}, K., Jul. 2011. {Can 21-cm observations discriminate
  between high-mass and low-mass galaxies as reionization sources?} ArXiv
  e-prints.

\bibitem[{{Jenkins} et~al.(2001){Jenkins}, {Frenk}, {White}, {Colberg}, {Cole},
  {Evrard}, {Couchman}, and {Yoshida}}]{Jenkins2001}
{Jenkins}, A., {Frenk}, C.~S., {White}, S.~D.~M., {Colberg}, J.~M., {Cole}, S.,
  {Evrard}, A.~E., {Couchman}, H.~M.~P., {Yoshida}, N., Feb. 2001. {The mass
  function of dark matter haloes}. \mnras 321, 372--384.

\bibitem[{{Jing}(1998)}]{Jing1998}
{Jing}, Y.~P., Aug. 1998. {Accurate Fitting Formula for the Two-Point
  Correlation Function of Dark Matter Halos}. \apjl 503, L9+.

\bibitem[{{Kauffmann} et~al.(1999){Kauffmann}, {Colberg}, {Diaferio}, and
  {White}}]{Kauffmann1999}
{Kauffmann}, G., {Colberg}, J.~M., {Diaferio}, A., {White}, S.~D.~M., Feb.
  1999. {Clustering of galaxies in a hierarchical universe - I. Methods and
  results at z=0}. \mnras 303, 188--206.

\bibitem[{{Kim} et~al.(2009){Kim}, {Park}, {Gott}, and {Dubinski}}]{Kim2009}
{Kim}, J., {Park}, C., {Gott}, III, J.~R., {Dubinski}, J., Aug. 2009. {The
  Horizon Run N-Body Simulation: Baryon Acoustic Oscillations and Topology of
  Large-scale Structure of the Universe}. \apj 701, 1547--1559.

\bibitem[{{Klypin} and {Holtzman}(1997)}]{Klypin.Holtzmann.1997}
{Klypin}, A., {Holtzman}, J., Dec. 1997. {Particle-Mesh code for cosmological
  simulations}. ArXiv Astrophysics e-prints.

\bibitem[{{Klypin} et~al.(1999){Klypin}, {Kravtsov}, {Valenzuela}, and
  {Prada}}]{Klypin1999}
{Klypin}, A., {Kravtsov}, A.~V., {Valenzuela}, O., {Prada}, F., Sep. 1999.
  {Where Are the Missing Galactic Satellites?} \apj 522, 82--92.

\bibitem[{{Klypin} et~al.(2010){Klypin}, {Trujillo-Gomez}, and
  {Primack}}]{Klypin.Trujillo-Gomez.Primack.2010}
{Klypin}, A., {Trujillo-Gomez}, S., {Primack}, J., Feb. 2010. {Halos and
  galaxies in the standard cosmological model: results from the Bolshoi
  simulation}. ArXiv e-prints.

\bibitem[{{Knollmann} and {Knebe}(2009)}]{Knollmann2009}
{Knollmann}, S.~R., {Knebe}, A., Jun. 2009. {AHF: Amiga's Halo Finder}. \apjs
  182, 608--624.

\bibitem[{{Kravtsov} et~al.(2004){Kravtsov}, {Berlind}, {Wechsler}, {Klypin},
  {Gottl{\"o}ber}, {Allgood}, and {Primack}}]{Kravtsov2004}
{Kravtsov}, A.~V., {Berlind}, A.~A., {Wechsler}, R.~H., {Klypin}, A.~A.,
  {Gottl{\"o}ber}, S., {Allgood}, B., {Primack}, J.~R., Jul. 2004. {The Dark
  Side of the Halo Occupation Distribution}. \apj 609, 35--49.

\bibitem[{{Kravtsov} and {Klypin}(1999)}]{Kravtsov1999}
{Kravtsov}, A.~V., {Klypin}, A.~A., Aug. 1999. {The Origin and Evolution of
  Halo Bias in Linear and Nonlinear Regimes}. \apj 520, 437--453.

\bibitem[{{Kravtsov} et~al.(1997){Kravtsov}, {Klypin}, and
  {Khokhlov}}]{Kravtsov.Klypin.Khokhlov.1997}
{Kravtsov}, A.~V., {Klypin}, A.~A., {Khokhlov}, A.~M., Jul. 1997. {Adaptive
  Refinement Tree: A New High-Resolution N-Body Code for Cosmological
  Simulations}. \apjs 111, 73--+.

\bibitem[{{Kuhlen} et~al.(2008){Kuhlen}, {Diemand}, and {Madau}}]{Kuhlen2008}
{Kuhlen}, M., {Diemand}, J., {Madau}, P., Oct. 2008. {The Dark Matter
  Annihilation Signal from Galactic Substructure: Predictions for GLAST}. \apj
  686, 262--278.

\bibitem[{{Lahav} et~al.(1991){Lahav}, {Lilje}, {Primack}, and
  {Rees}}]{1991MNRAS.251..128L}
{Lahav}, O., {Lilje}, P.~B., {Primack}, J.~R., {Rees}, M.~J., Jul. 1991.
  {Dynamical effects of the cosmological constant}. \mnras 251, 128--136.

\bibitem[{{Lemson} et~al.(2011){Lemson}, {Budavari}, and {Szalay}}]{Lemson2011}
{Lemson}, G., {Budavari}, T., {Szalay}, A., 2011. {Implementing a General
  Spatial Indexing Library for Relational Databases of Large Numerical
  Simulations}. Accepted for publication in SSDBM 2011.

\bibitem[{{Lemson} and {Springel}(2006)}]{2006ADASS}
{Lemson}, G., {Springel}, V., 2006. {Cosmological Simulations in a Relational
  Database: Modelling and Storing Merger Trees}. ADASS.

\bibitem[{{Lemson} and {Virgo Consortium}(2006)}]{2006astro.ph..8019L}
{Lemson}, G., {Virgo Consortium}, Aug. 2006. {Halo and Galaxy Formation
  Histories from the Millennium Simulation: Public release of a VO-oriented and
  SQL-queryable database for studying the evolution of galaxies in the
  LambdaCDM cosmogony}. ArXiv Astrophysics e-prints.

\bibitem[{{Macci{\`o}} et~al.(2008){Macci{\`o}}, {Dutton}, and {van den
  Bosch}}]{Maccio2008}
{Macci{\`o}}, A.~V., {Dutton}, A.~A., {van den Bosch}, F.~C., Dec. 2008.
  {Concentration, spin and shape of dark matter haloes as a function of the
  cosmological model: WMAP1, WMAP3 and WMAP5 results}. \mnras 391, 1940--1954.

\bibitem[{{Moore} et~al.(1999){Moore}, {Ghigna}, {Governato}, {Lake}, {Quinn},
  {Stadel}, and {Tozzi}}]{Moore1999}
{Moore}, B., {Ghigna}, S., {Governato}, F., {Lake}, G., {Quinn}, T., {Stadel},
  J., {Tozzi}, P., Oct. 1999. {Dark Matter Substructure within Galactic Halos}.
  \apjl 524, L19--L22.

\bibitem[{{More} et~al.(2011){More}, {Kravtsov}, {Dalal}, and
  {Gottl{\"o}ber}}]{More.et.al.2011}
{More}, S., {Kravtsov}, A., {Dalal}, N., {Gottl{\"o}ber}, S., Feb. 2011. {The
  overdensity and masses of the friends-of-friends halos and universality of
  the halo mass function}. ArXiv e-prints.

\bibitem[{{MSDN}(2008)}]{TSQL}
{MSDN}, 2008. {Transact-SQL Reference}.
\newline\urlprefix\url{http://msdn.microsoft.com/en-us/library/bb510741.aspx}

\bibitem[{{Navarro} et~al.(1997){Navarro}, {Frenk}, and {White}}]{NFW1997}
{Navarro}, J.~F., {Frenk}, C.~S., {White}, S.~D.~M., Dec. 1997. {A Universal
  Density Profile from Hierarchical Clustering}. \apj 490, 493--+.

\bibitem[{{Neto} et~al.(2007){Neto}, {Gao}, {Bett}, {Cole}, {Navarro}, {Frenk},
  {White}, {Springel}, and {Jenkins}}]{Neto2007}
{Neto}, A.~F., {Gao}, L., {Bett}, P., {Cole}, S., {Navarro}, J.~F., {Frenk},
  C.~S., {White}, S.~D.~M., {Springel}, V., {Jenkins}, A., Nov. 2007. {The
  statistics of {$\Lambda$} CDM halo concentrations}. \mnras 381, 1450--1462.

\bibitem[{{Peebles}(1970)}]{Peebles1970}
{Peebles}, P.~J.~E., Feb. 1970. {Structure of the Coma Cluster of Galaxies}.
  \aj 75, 13.

\bibitem[{{Prada} et~al.(2011){Prada}, {Klypin}, {Cuesta}, {Betancort-Rijo},
  and {Primack}}]{Prada2011}
{Prada}, F., {Klypin}, A., {Cuesta}, A., {Betancort-Rijo}, J., {Primack}, J.,
  Apr. 2011. {Halo concentrations in the standard LCDM Cosmology}.

\bibitem[{{Prada} et~al.(2006){Prada}, {Klypin}, {Simonneau}, {Betancort-Rijo},
  {Patiri}, {Gottl{\"o}ber}, and {Sanchez-Conde}}]{2006ApJ...645.1001P}
{Prada}, F., {Klypin}, A.~A., {Simonneau}, E., {Betancort-Rijo}, J., {Patiri},
  S., {Gottl{\"o}ber}, S., {Sanchez-Conde}, M.~A., Jul. 2006. {How Far Do They
  Go? The Outer Structure of Galactic Dark Matter Halos}. \apj 645, 1001.

\bibitem[{{Sheth} and {Tormen}(2002)}]{Sheth2002}
{Sheth}, R.~K., {Tormen}, G., Jan. 2002. {An excursion set model of
  hierarchical clustering: ellipsoidal collapse and the moving barrier}. \mnras
  329, 61--75.

\bibitem[{{skyserver.sdss.org}(2008)}]{SDSS_Database}
{skyserver.sdss.org}, 2008. \url{http://skyserver.sdss.org}.

\bibitem[{{Somerville} et~al.(2011){Somerville}, {Gilmore}, {Primack}, and
  {Dominguez}}]{Somerville2011}
{Somerville}, R.~S., {Gilmore}, R.~C., {Primack}, J.~R., {Dominguez}, A., Apr.
  2011. {Galaxy Properties from the Ultra-violet to the Far-Infrared:
  Lambda-CDM models confront observations}. ArXiv e-prints.

\bibitem[{{Somerville} and {Primack}(1999)}]{Somerville1999}
{Somerville}, R.~S., {Primack}, J.~R., Dec. 1999. {Semi-analytic modelling of
  galaxy formation: the local Universe}. \mnras 310, 1087--1110.

\bibitem[{{Springel}(2005)}]{Springel.2005}
{Springel}, V., Dec. 2005. {The cosmological simulation code GADGET-2}. \mnras
  364, 1105--1134.

\bibitem[{{Springel} et~al.(2008){Springel}, {Wang}, {Vogelsberger}, {Ludlow},
  {Jenkins}, {Helmi}, {Navarro}, {Frenk}, and {White}}]{Springel2008}
{Springel}, V., {Wang}, J., {Vogelsberger}, M., {Ludlow}, A., {Jenkins}, A.,
  {Helmi}, A., {Navarro}, J.~F., {Frenk}, C.~S., {White}, S.~D.~M., Dec. 2008.
  {The Aquarius Project: the subhaloes of galactic haloes}. \mnras 391,
  1685--1711.

\bibitem[{Springel et~al.(2005)Springel, White, Jenkins, Frenk, Yoshida, Gao,
  Navarro, Thacker, Croton, Helly, Peacock, Cole, Thomas, Couchman, Evrard,
  Colberg, and Pearce}]{millennium}
Springel, V., White, S. D.~M., Jenkins, A., Frenk, C.~S., Yoshida, N., Gao, L.,
  Navarro, J., Thacker, R., Croton, D., Helly, J., Peacock, J.~A., Cole, S.,
  Thomas, P., Couchman, H., Evrard, A., Colberg, J., Pearce, F., Jun. 2005.
  {Simulating the joint evolution of quasars, galaxies and their large-scale
  distribution}. Nature 435, 629--636.

\bibitem[{{Stadel} et~al.(2009){Stadel}, {Potter}, {Moore}, {Diemand}, {Madau},
  {Zemp}, {Kuhlen}, and {Quilis}}]{Stadel2009}
{Stadel}, J., {Potter}, D., {Moore}, B., {Diemand}, J., {Madau}, P., {Zemp},
  M., {Kuhlen}, M., {Quilis}, V., Sep. 2009. {Quantifying the heart of darkness
  with GHALO - a multibillion particle simulation of a galactic halo}. \mnras
  398, L21--L25.

\bibitem[{{Teyssier} et~al.(2009){Teyssier}, {Pires}, {Prunet}, {Aubert},
  {Pichon}, {Amara}, {Benabed}, {Colombi}, {Refregier}, and
  {Starck}}]{Teyssier2009}
{Teyssier}, R., {Pires}, S., {Prunet}, S., {Aubert}, D., {Pichon}, C., {Amara},
  A., {Benabed}, K., {Colombi}, S., {Refregier}, A., {Starck}, J.-L., Apr.
  2009. {Full-sky weak-lensing simulation with 70 billion particles}. \aap 497,
  335--341.

\bibitem[{{Tinker} et~al.(2008){Tinker}, {Kravtsov}, {Klypin}, {Abazajian},
  {Warren}, {Yepes}, {Gottl{\"o}ber}, and {Holz}}]{Tinker2008}
{Tinker}, J., {Kravtsov}, A.~V., {Klypin}, A., {Abazajian}, K., {Warren}, M.,
  {Yepes}, G., {Gottl{\"o}ber}, S., {Holz}, D.~E., Dec. 2008. {Toward a Halo
  Mass Function for Precision Cosmology: The Limits of Universality}. \apj 688,
  709--728.

\bibitem[{{Tinker} et~al.(2010){Tinker}, {Robertson}, {Kravtsov}, {Klypin},
  {Warren}, {Yepes}, and {Gottl{\"o}ber}}]{Tinker2010}
{Tinker}, J.~L., {Robertson}, B.~E., {Kravtsov}, A.~V., {Klypin}, A., {Warren},
  M.~S., {Yepes}, G., {Gottl{\"o}ber}, S., Dec. 2010. {The Large-scale Bias of
  Dark Matter Halos: Numerical Calibration and Model Tests}. \apj 724,
  878--886.

\bibitem[{{Trujillo-Gomez} et~al.(2010){Trujillo-Gomez}, {Klypin}, {Primack},
  and {Romanowsky}}]{Trujillo2010}
{Trujillo-Gomez}, S., {Klypin}, A., {Primack}, J., {Romanowsky}, A., Mar. 2010.
  {LCDM Correctly Predicts Basic Statistics of Galaxies: Luminosity-Velocity
  Relation, Baryonic Mass-Velocity Relation, and Velocity Function}. ArXiv
  e-prints.

\bibitem[{{Vale} and {Ostriker}(2004)}]{Vale2004}
{Vale}, A., {Ostriker}, J.~P., Sep. 2004. {Linking halo mass to galaxy
  luminosity}. \mnras 353, 189--200.

\bibitem[{{van den Bosch} et~al.(2007){van den Bosch}, {Yang}, {Mo},
  {Weinmann}, {Macci{\`o}}, {More}, {Cacciato}, {Skibba}, and
  {Kang}}]{Bosch2007}
{van den Bosch}, F.~C., {Yang}, X., {Mo}, H.~J., {Weinmann}, S.~M.,
  {Macci{\`o}}, A.~V., {More}, S., {Cacciato}, M., {Skibba}, R., {Kang}, X.,
  Apr. 2007. {Towards a concordant model of halo occupation statistics}. \mnras
  376, 841--860.

\bibitem[{{Warren} et~al.(2006){Warren}, {Abazajian}, {Holz}, and
  {Teodoro}}]{Warren2006}
{Warren}, M.~S., {Abazajian}, K., {Holz}, D.~E., {Teodoro}, L., Aug. 2006.
  {Precision Determination of the Mass Function of Dark Matter Halos}. \apj
  646, 881--885.

\bibitem[{{Wechsler} et~al.(2006){Wechsler}, {Zentner}, {Bullock}, {Kravtsov},
  and {Allgood}}]{Wechsler2006}
{Wechsler}, R.~H., {Zentner}, A.~R., {Bullock}, J.~S., {Kravtsov}, A.~V.,
  {Allgood}, B., Nov. 2006. {The Dependence of Halo Clustering on Halo
  Formation History, Concentration, and Occupation}. \apj 652, 71--84.

\bibitem[{Wetzel and White(2010)}]{Wetzel2010}
Wetzel, A.~R., White, M., 2010. {What determines satellite galaxy disruption?}
  MNRAS 403, 1072--1088.

\bibitem[{{White}(1976)}]{White1976}
{White}, S.~D.~M., Dec. 1976. {The dynamics of rich clusters of galaxies}.
  \mnras 177, 717.

\bibitem[{{Zentner} et~al.(2005){Zentner}, {Berlind}, {Bullock}, {Kravtsov},
  and {Wechsler}}]{Zentner2005}
{Zentner}, A.~R., {Berlind}, A.~A., {Bullock}, J.~S., {Kravtsov}, A.~V.,
  {Wechsler}, R.~H., May 2005. {The Physics of Galaxy Clustering. I. A Model
  for Subhalo Populations}. \apj 624, 505--525.

\bibitem[{{Zhao} et~al.(2003){Zhao}, {Mo}, {Jing}, and {B{\"o}rner}}]{Zhao2003}
{Zhao}, D.~H., {Mo}, H.~J., {Jing}, Y.~P., {B{\"o}rner}, G., Feb. 2003. {The
  growth and structure of dark matter haloes}. \mnras 339, 12--24.

\end{thebibliography}

\end{document}